\documentclass[12pt,preprint,apj]{emulateapj}
\newcommand{\kms}{{\rm km~s^{-1}}}

\newcommand{\rvir}{R_{200}}

\newcommand{\efrac}{f_{\rm E}}
\newcommand{\nbg}{N_{\rm nei}}

\usepackage{color}

\shorttitle{Galaxy Evolution in Compact Groups}
\shortauthors{Lee, G.-H., et al.}

\begin{document}


\title{The Fastest Galaxy Evolution in an Unbiased Compact Group Sample with \textit{WISE}}

\author{Gwang-Ho Lee$^{1}$, Ho Seong Hwang$^{2}$, Jubee Sohn$^{3}$, and Myung Gyoon Lee$^{1}$} 
\affil{$^{1}$Department of Physics and Astronomy, Seoul National University, 
1 Gwanak-ro, Gwanak-gu, Seoul 08826, Republic of Korea}
\affil{$^{2}$School of Physics, Korea Institute for Advanced Study,
85 Hoegiro, Dongdaemun-gu, Seoul 02455, Republic of Korea}
\affil{$^{3}$Smithsonian Astrophysical Observatory, 60 Garden Street, Cambridge, MA 02138, USA}
\email{ghlee@astro.snu.ac.kr, mglee@astro.snu.ac.kr}


\begin{abstract}

We study the mid-infrared (MIR) properties of galaxies 
in compact groups and their environmental dependence 
using the \textit{Wide-field Infrared Survey Explorer (WISE)} data.
We use a volume-limited sample of 670 compact groups and 
their 2175 member galaxies with $M_r< -19.77$ and $0.01<z<0.0741$, 
drawn from \citet{sohn+16}, which were identified using a friends-of-friends algorithm.
Among the 2175 galaxies, 1541 galaxies are detected at 
\textit{WISE} 12 $\micron$ with a signal-to-noise ratio greater than 3.
Among the 1541 galaxies, 433 AGN-host galaxies are identified by using both 
optical and MIR classification scheme.
Using the remaining 1108 non-AGN galaxies, we find that the MIR $[3.4]-[12]$ colors of compact group early-type galaxies 
are on average bluer than those of cluster early-type galaxies.
When compact groups have both early- and late-type member galaxies, 
the MIR colors of the late-type members in those compact groups
are bluer than the MIR colors of cluster late-type galaxies.
As compact groups are located in denser regions,
they tend to have larger early-type galaxy fractions
and bluer MIR color galaxies. 
These trends are also seen for neighboring galaxies around compact groups.
However, compact group member galaxies always have larger 
early-type galaxy fractions and bluer MIR colors than their neighboring galaxies.
Our findings suggest that the properties of compact group galaxies depend on 
both internal and external environments of compact groups, 
and that galaxy evolution is faster in compact groups than in the central regions of clusters.

\end{abstract}

\keywords{galaxies: evolution -- galaxies: groups: general -- galaxies: interactions}

\section{Introduction}

Compact groups of galaxies contain several galaxies within just a few
tens of kilo-parsec scales.
This makes compact groups have extremely high
galaxy number densities (the median galaxy number density of 
log$(\rho /$[$h^{-3}$ Mpc$^3$])$=4.36$, \citealp{sohn+16}), higher than 
the galaxy number densities of cluster and group environments 
(the median galaxy number density of 
log$(\rho /$[$h^{-3}$ Mpc$^3$])$=1.97$ for clusters and groups in the 
A2199 supercluster, \citealp{lee+15,sohn+15}).
The velocity dispersions of compact groups 
are much smaller \citep[$< 879~\kms$ 
with a median value of $194~\kms$ with a standard deviation of 
$156~\kms$,][Table 7]{sohn+16} than 
those of clusters \citep[$\sim$400$-1300~\kms$ with a median value of $778~\kms$ and a standard deviation of $208~\kms$,][]{rines+13}.
These two characteristics make compact groups 
an ideal environment for frequent interactions 
and mergers between galaxies \citep[e.g.,][]{rubin+90,rodrigue+95,amram+07,coziol+07,plauchu+10,gallagher+10,konstantopoulos+12,sohn+13,vogt+15}. 

Numerical simulations suggested that galaxies in a compact group
should merge into a single elliptical galaxy within a few Gyrs, 
causing the compact group to disappear \citep{barnes85,barnes89,mamon87}. 
\citet{gallagher+10} also reached a similar conclusion through their study 
of star cluster age-dating and neutral hydrogen content in the Hickson compact 
group 31; the compact group should merge into a single elliptical within 1 Gyr
(see also \citealt{rubin+90}).
\citet{kroupa15} suggested that the abundance of compact groups should 
decrease as the redshift decreases from 
$z\simeq0.1$ to $z=0$ in the $\Lambda$CDM
universe. However, \citet{sohn+15} found that the abundance of 
compact groups changes little with redshifts at $0.01<z<0.22$, using 
a spectroscopically complete sample of 332 compact groups.  
\citet{kroupa15} also suggested an alternative scenario that 
compact groups do not merge, which predicts a constant abundance of 
compact groups at $z<0.1$. However, this alternative scenario is based on 
the assumption of a universe without dark matter.

\citet{diaferio+94} suggested that compact groups 
replenish themselves with new members from the surrounding environment, 
thereby extending their lifetimes to the current epoch. 
The finding of \citet{sohn+15}, constant abundance of compact group 
at $0.01<z<0.22$, can be explained by this replenishment model.
This replenishment model is supported by several observational findings 
that showed that many ($>50\%$) compact groups are embedded in or associated with 
larger-scale structures 
\citep{rood+94,ramella+94,decarvalho+94,ribeiro+98,mendel+11,pompei+12}. 
However, \citet{diaz+15} showed that only 27\% of compact groups 
are embedded in surrounding structures.

Unfortunately, many existing compact group catalogs suffer 
from some observational biases, 
which make them difficult to use for testing compact group formation models 
and for studying the relation between compact groups and their surrounding 
environments. 
For example, \citet{mcconnachie+09} constructed a large sample of 
compact groups, from the photometric data ($m_r<21$) of 
Sloan Digital Sky Survey (SDSS) Data Release 6 \citep[DR6,][]{adelman+08}, 
including 77,088 compact group candidates and their 313,508 tentative member galaxies. 
Their sample is the largest one currently, but it is highly 
contaminated with interlopers \citep[$>50\%$,][]{mendel+11,sohn+13} 
because the sample is based only on photometric information using
Hickson's selection criteria \citep{hickson82}.
To reduce this contamination, \citet{sohn+15} compiled the redshifts from 
Fred Lawrence Whipple Observatory (FLWO)/FAST observations and 
from the SDSS DR12 \citep{alam+15} 
for the photometric sample of compact group galaxies 
in \citet{mcconnachie+09}, and constructed a catalog of 332 compact groups.
This catalog is currently the largest one with complete spectroscopic redshifts.
However, this sample is affected by an isolation criterion
from Hickson's selection criteria.
The isolation criterion requires that no galaxies be present in a radius of up to 
three times the group radius.
This criterion was adopted to avoid selecting very dense regions
like cluster cores, 
but this could introduce a bias that misses nearby compact groups 
with apparently large sizes and embedded systems in dense environments 
\citep{barton+96}.

Recently, \citet{sohn+16} published a new catalog of 1588 compact groups 
and their 5178 member galaxies at $0.01<z<0.19$ using  
SDSS DR12 spectroscopic data supplemented by additional redshifts 
from the literature and from FLWO/FAST observations.
They applied a friends-of-friends algorithm to identify the compact groups 
without using the isolation criterion.
As a result, their catalog successfully contains nearby ($z<0.05$) compact groups 
and embedded systems in dense regions. 
This is important in order to study the relation between compact groups and 
their surrounding environments without any sample bias.

Previous observational studies suggested that galaxy evolution in compact groups 
is faster than in the field from the comparisons of Lick indices 
\citep{proctor+04,mendes+05,de_la_rosa+07}, 
the star formation quenching timescales \citep{plauchu+12}, 
specific star formation rates (SFRs) distribution 
\citep{tzanavaris+10,coenda+15,lenkic+16},
and color distribution \citep{bitsakis+10,bitsakis+11,bitsakis+16,walker+10,walker+12}.
\citet{proctor+04} and \citet{mendes+05} found that 
the stellar ages and the early-type galaxy fractions of compact group galaxies 
are similar to those of cluster galaxies, which suggests that 
compact group galaxies evolve as fast as cluster galaxies.

On the other hand, \citet{johnson+07} and \citet{walker+10, walker+12} 
showed that compact group galaxies exhibit a bimodal distribution 
in the \textit{Spitzer} IRAC ($3.6-8.0$ $\micron$) color space 
with a statistically evident gap, so-called ``canyon'', 
between star-forming galaxies with MIR red colors 
and quiescent galaxies with MIR blue colors. 
Recently, \citet{zucker+16} newly identified the canyon galaxies in the 
\textit{WISE} color-color diagram.
This gap may suggest the accelerated galaxy evolution in compact groups 
because the gap is not seen in comparison galaxy samples from 
field, from interacting pairs, and from the center of the Coma cluster 
\citep{walker+10,walker+12,walker+13}. 
\citet{cluver+13} found that compact group galaxies in the gap 
mostly show enhanced warm H$_2$ emission from the observations with 
the \textit{Spitzer} Infrared Spectrograph, which could be caused by shock 
heating \citep{appleton+06,cluver+10}. This result implies that shock heating 
may be responsible for rapid evolution of galaxies in compact groups.

However, these results are mainly based on the catalogs of compact groups 
selected using Hickson's criteria, which can introduce a sample bias.
We therefore revisit the issues on the comparison of compact groups galaxies 
with cluster and field galaxies using the unbiased sample of compact groups 
from \citet{sohn+16}.
In this paper, we use the \textit{Wide-field Infrared Survey Explorer}
\citep[\textit{WISE},][]{wright+10} mid-infrared (MIR) data to study 
the environmental effects on compact group galaxies and the relation between 
compact groups and their surrounding environments.
The MIR data are useful indicators of mean stellar ages \citep{piovan+03,ko+09}, 
especially for compact group galaxies that are mainly dominated by old stellar 
populations. Red-sequence galaxies with small amounts of young ($<1$ Gyr) and 
intermediate-age (1-10 Gyr) stellar populations can be distinguished from 
those without the populations in MIR color space \citep{ko+13,ko+16,lee+15}. 
The sample we use in this paper contains 670 compact groups 
with 2175 member galaxies drawn from the catalog of \citet{sohn+16}.
When we investigate the MIR properties of galaxies in Sections 3 and 4, 
we use 1541 member galaxies detected at $12~\micron$ with 
a signal-to-noise ratio (S/N) greater than 3.
This sample is larger than those of previous MIR studies.
For example,
the samples of \citet{walker+12} and several previous studies 
\citep{johnson+07,bitsakis+10,bitsakis+11,walker+10} have less than 50 compact 
groups, the sample of \citet{walker+13} has 99 compact groups and 
348 member galaxies, and 
the sample of \citet{zucker+16} has 163 compact groups and 567 member galaxies.

Section \ref{data} describes the compact group sample and 
comparison samples of cluster/field galaxies. 
In Section 3 we compare the properties of galaxies in compact groups with 
those in other environments, and investigate how environment affects 
the properties of galaxies in compact groups. 
In Section 4 we discuss the environmental effects on galaxy evolution 
in compact groups
and the relation between compact groups and their surrounding environments.
In Section 5 we summarize our results and present conclusions.
Throughout, we adopt flat $\Lambda$CDM cosmological parameters: 
$H_0=70$ km s$^{-1}$ Mpc$^{-1}$, $\Omega_{\Lambda}=0.7$, and 
$\Omega_{m}=0.3$.

\section{Data}\label{data}


\begin{deluxetable*}{cccccccc}
\tablecolumns{8}
\tablewidth{0pc}
\tablecaption{SDSS-related Physical Parameters of Compact Group Galaxies}
\tablehead{
\colhead{Host Group$^a$} & \colhead{Galaxy ID} & \colhead{R.A.} & \colhead{Decl} & \colhead{Redshift} & 
\colhead{$M_r^b$} & \colhead{$u-r^c$} & \colhead{Morph$^d$} 
}
\startdata
V1CG001 & 1237648705657307198  & 198.233322 &   1.007515 & $0.073839\pm0.000029$ &  $-19.82\pm2.21$ & $2.28\pm0.10$ &  2 \\
V1CG001 & 1237648705657307347  & 198.229294 &   1.010990 & $0.072679\pm0.000020$ &  $-20.11\pm0.01$ & $2.57\pm0.06$ &  1 \\
V1CG001 & 1237648705657307315  & 198.218872 &   1.019821 & $0.070435\pm0.000017$ &  $-21.16\pm0.01$ & $2.67\pm0.04$ &  1 \\
V1CG002 & 1237661126155436166  & 139.939529 & 33.745014 & $0.017118\pm0.000007$ &  $-21.22\pm0.06$ & $2.77\pm0.01$ &  1 \\
V1CG002 & 1237661126155436169  & 139.922531 & 33.738174 & $0.020334\pm0.000013$ &  $-20.72\pm0.05$ & $2.61\pm0.01$ &  2 \\
V1CG002 & 1237661126155436164  & 139.945221 & 33.749741 & $0.023006\pm0.000003$ &  $-22.84\pm0.01$ & $2.91\pm0.01$ &  1 \\
V1CG003 & 1237661136886431890  & 154.733276 & 37.285831 & $0.048809\pm0.000019$ &  $-20.41\pm0.01$ & $1.91\pm0.03$ &  2 \\
V1CG003 & 1237661136886497282  & 154.743576 & 37.300205 & $0.047580\pm0.000007$ &  $-20.84\pm0.01$ & $1.68\pm0.01$ &  2 \\
V1CG003 & 1237661136886497353  & 154.747711 & 37.308155 & $0.047489\pm0.000008$ &  $-20.08\pm0.01$ & $1.99\pm0.02$ &  2
\enddata
\label{tab1}
\tablecomments{The full table is available in the online journal. 
A portion is shown here for guidance regarding its form and content.\\
$^a$ Group ID from Table 5 of \citet{sohn+16}.\\
$^b$ The $k_{z=0.1}$-, and evolution-corrected $r$-band Petrosian absolute magnitudes.\\
$^c$ The SDSS extinction- and $k$-corrected model magnitudes. \\
$^d$ Galaxy morphology. 1 indicates an early-type galaxy, while 2 indicates a late-type galaxies. \\
}
\end{deluxetable*}

\citet{sohn+16} constructed a catalog of compact groups using 
the spectroscopic sample of SDSS DR12 galaxies with $m_r < 17.77$. 
The completeness of the spectroscopic data in SDSS is low for bright galaxies 
at $m_r < 14.5$ because of saturation and cross-talk in 
the spectrograph, and for the galaxies in high-density regions 
(e.g., galaxy clusters) because of the fiber collision (see Figure 1 of \citealt{park+09}).
Therefore, \citet{sohn+16} supplemented the catalog with redshifts 
from the literature
(\citealt{hill+93,hill+98,wegner+96,wegner+99,slinglend+98,falco+99}; see also 
\citealt{hwang+10}) and from FAST observations at FLWO \citep{sohn+15}. 
They applied the friends-of-friends algorithm 
with a projected linking length of $\Delta D = 50~h^{-1}$ kpc and 
a radial linking length of $| \Delta V | = 1000$ km s$^{-1}$, 
and constructed a magnitude-limited ($m_r<17.77$) sample of 1588 compact 
groups with each consisting of three or more member galaxies at $0.01<z<0.19$. 
This new catalog contains 18 times as many systems and reaches three times 
the depth of the catalog of \citet{barton+96}, which is also based on 
the friends-of-friends algorithm.

\citet{sohn+16} also constructed two volume-limited 
subsamples: the V1 sample of galaxies with $M_r<-19.77$ and 
$0.01<z<0.0741$ and the V2 sample of galaxies with $M_r<-20.77$ 
and $0.01<z<0.1154$ (V2)\footnote{
\citet{sohn+16} used $M_r<-19+5{\rm log}h$ 
and $M_r<-20+5{\rm log}h$ with 
$h=1$ (i.e., $H_0=100~\kms~{\rm Mpc}^{-1}$). However, we adopt 
$h=0.7$ (i.e., $H_0=70~\kms~{\rm Mpc}^{-1}$), 
which results in $-19+5{\rm log}h\simeq-19.77$ and 
$-20+5{\rm log}h\simeq-20.77$.}.
These volume-limited samples contain 670 and 297 compact groups 
that are independently identified through the friends-of-friends algorithm. 
Unlike systems in the magnitude-limited sample, the volume-limited samples
have systems with a median stellar mass independent of redshift. 
Therefore, the volume-limited samples allow us to study the properties of 
compact groups without any sample bias that could be introduced 
in the magnitude-limited sample.
Details for the catalogs and the compact group selection 
are described in Sections 2 and 3 of \citet{sohn+16}.

In this study, we use the V1 sample of 670 compact groups and 
their 2175 member galaxies with $M_r < -19.77$ and $0.01<z<0.0741$. 
To study the MIR properties of these galaxies, we cross-correlate the galaxies 
with the objects in the \textit{ALLWISE} source 
catalog\footnote{http://wise2.ipac.caltech.edu/docs/release/allwise/}
using a matching tolerance of 3$^{\prime\prime}$, corresponding to 
about half of the FWHM of the PSF at 3.4 $\micron$.
Among the 2175 galaxies, 2067 (95\%) are matched with \textit{ALLWISE} sources. 
We use the profile-fitting magnitudes of the sources at four MIR bands 
(3.4 $\micron$, 4.6 $\micron$, 12 $\micron$, and 22 $\micron$).
When we investigate the MIR properties of galaxies, we use only galaxies 
detected at 12 $\micron$ with ${\rm S/N}>3$. 
Because the \textit{ALLWISE} magnitudes with ${\rm S/N}<2$ are just upper limits, 
we choose only the galaxies with ${\rm S/N}>3$ to safely use the magnitudes.
Among the 2175 galaxies, 1541 galaxies (71\%) are detected at 12 $\micron$. 
If we adopt the S/N cut of 5, the sample size is reduced from 1541 to 863 galaxies
by 44\%. However, we confirm that our conclusions do not change much
even if we use this small sample.

To compare the physical properties of compact group galaxies with those 
in other environments, we use two comparison samples of cluster and field galaxies. 
\citet{hwang+12a} constructed a sample of 129 relaxed Abell clusters at 
$0.02<z<0.14$ using the spectroscopic sample of the SDSS DR7 
\citep{abazajian+09}. 
The cluster sample also includes the galaxies at $R<10\rvir$ 
($\rvir$ is the virial radius of the cluster) and 
$\Delta V = | V_{\rm gal} - V_{\rm cl} | < 1000$ km s$^{-1}$ 
($V_{\rm gal}$ and $V_{\rm cl}$ are radial velocity of a galaxy and 
systematic velocity of the cluster).
From this sample, we selected 2433 galaxies at $R<\rvir$ 
as the cluster galaxy sample, and 6312 galaxies at $5\rvir<R<10\rvir$ 
as the field galaxy sample. 
\citet{park+09} showed that the fraction of early-type galaxies 
decreases with increasing clustercentric radius at $R<3\rvir$. 
However, the early-type galaxy fraction is nearly constant at $R>4\rvir$ (see their Figure 4), suggesting that the galaxies at $5\rvir<R<10\rvir$ can be considered to be field galaxies.
The galaxies in the two comparison samples also satisfy the criterion of 
$M_r<-19.77$ and $0.01<z<0.0741$ as do those in the V1 sample.

The optical parameters of galaxies that we consider in this study are 
$r$-band Petrosian absolute magnitude ($M_r$), $u-r$ color, and morphology. 
The $M_r$ was computed using the Galactic reddening correction \citep{schlegel+98}, 
$K$-corrections \citep{blanton+07}, shifted to $z=0.1$, and 
evolution correction, $E(z)=1.6(z-0.1)$ \citep{tegmark+04}.
The $u-r$ color was computed using extinction- and $K$-corrected (to $z=0.1$)
$u$- and $r$-band model magnitudes.
Galaxy morphology data are mainly from the Korea Institute for Advanced Study
(KIAS) DR7 value-added galaxy catalog \citep{choi+10}. 
Galaxies in this catalog are morphologically classified into early (E/S0) and 
late types (S/Irr) based on an automated scheme with $u-r$ color, 
the $g-i$ color gradient, and the $i$-band concentration index \citep{park+05}. 
For galaxies not included in the KIAS DR7 value-added galaxy catalog, \citet{sohn+16} 
classified them into early and late types through visual inspection. 
In Table \ref{tab1}, we list the group ID, galaxy ID, R.A., declination, redshift, 
$M_r$, $u-r$, and morphology of the 2175 compact group galaxies.

\begin{figure}
\centering
\includegraphics[scale=0.7]{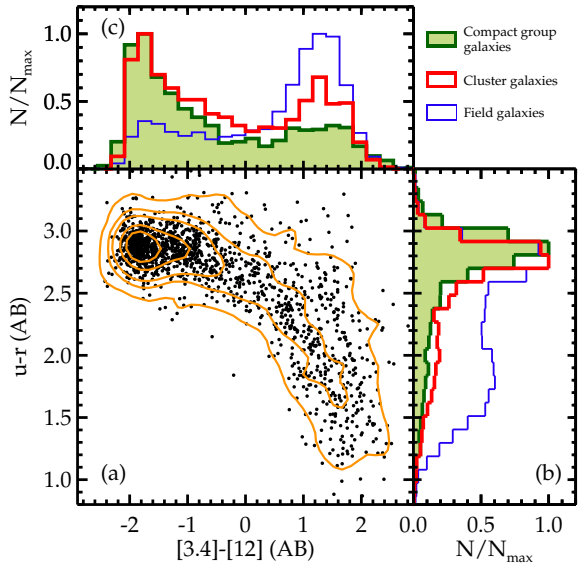}
\caption{(a) $u-r$ versus $[3.4]-[12]$ diagram for the 1541 compact group galaxies 
detected at 12 $\micron$. Contours represent the number density distribution.
In panels (b) and (c), we display the $u-r$ and $[3.4]-[12]$ distribution for 
the compact group galaxies with the filled histograms.  For comparison, we also plot 
the MIR color distribution of the 1461 cluster galaxies (open histogram with thick lines) 
and for the 5084 field galaxies (open histogram with thin lines).
}
\label{colorhist}
\end{figure}

\section{Results}\label{result} 

\subsection{Galaxy Color Distributions}

Figure \ref{colorhist} shows the $u-r$ and $[3.4]-[12]$ distributions 
of the 1541 compact group galaxies detected at 12 $\micron$. 
The mean $u-r$ error is 0.07, while the mean $[3.4]-[12]$ error is 0.12. 
In this diagram, compact group galaxies are distributed from the top-left corner
to the bottom-right corner because galaxies with bluer optical colors tend to have 
redder MIR colors.

In panels (b) and (c), we plot the $u-r$ and $[3.4]-[12]$ distributions 
of compact group galaxies.
For comparison, we also plot the distributions of cluster and field galaxies.
As shown in panel (b), 
$u-r$ histograms for the three galaxy samples peak at $u-r\simeq2.6$. 
However, 
the fraction of galaxies with $u-r>2.5$ in the compact group galaxy sample is 
$73.8\%\pm1.1\%$, which is larger than that of the cluster galaxy sample 
($64.1\%\pm1.2\%$) and of the field galaxy sample ($38.4\%\pm0.7\%$).
The mean $u-r$ is $2.65\pm0.02$ in the compact group galaxy sample, 
which is also larger  than the $2.50\pm0.01$ of the cluster galaxy sample 
and the $2.22\pm0.01$ of the field galaxy sample\footnote{
The errors in the fractions and in the mean values are the standard deviations 
derived from 1000-times resamplings.}.
These suggest that compact group galaxies are dominated by optical 
red-sequence galaxies much like cluster galaxies, which is consistent with 
the findings of \citet{walker+13} and \citet{zucker+16}. 

Optical red-sequence galaxies, located in a narrow $u-r$ range of $2.5-3.2$,
have MIR colors ranging from $[3.4]-[12]\simeq-2.4$ to $[3.4]-[12]\simeq0$. 
\citet{lee+15} divided optical red-sequence galaxies into MIR blue galaxies 
(i.e., $[3.4]-[12]<-1.55$) and MIR green galaxies (i.e., $-1.55\leq[3.4]-[12]<-0.3$)\footnote{This criterion is determined from the decomposition 
of the $[3.4]-[12]$ color distribution with three Gaussian functions 
using the Gaussian mixture modeling 
\citep[][see Figure 2 in \citealt{lee+15}]{muratov+10}. 
We stress that this MIR green color selection criterion 
is not identical to the ``MIR green valley'' one in Section 3.2.}. 
The MIR blue galaxies are composed of stellar populations older than 10 Gyr, 
while the MIR green galaxies have small ($\sim$5\%) mass fractions of 
relatively young ($0.5-1$ Gyr) or intermediate ($1-10$ Gyr) stellar populations 
\citep{piovan+03,ko+13,lee+15,ko+16}. Optical colors trace star formation on 
timescales up to 1 Gyr \citep{schawinski+14}. Therefore, the MIR green  
galaxies are not distinguishable from the MIR blue galaxies in optical colors
because both lie in the optical red-sequence \citep{walker+12,walker+13,ko+13,lee+15,zucker+16}. Thus, we need to use MIR colors instead of optical colors 
to investigate whether galaxies in compact groups experienced 
different star formation histories from those in clusters, or not.

In panel (c) of Figure \ref{colorhist}, we compare the MIR color distributions between 
the three different galaxy samples. The cluster galaxy sample shows a blue peak 
at $[3.4]-[12]\simeq-1.7$ and a red peak $[3.4]-[12]\simeq1.3$. 
The blue peak is higher than the red peak. 
The compact group galaxy sample shows a strong blue peak and a red bump, 
but the field galaxy sample shows a different color distribution with a blue bump 
and a strong red peak.
The mean $[3.4]-[12]$ is $-0.54\pm0.03$ for the compact group galaxy sample, 
$-0.29\pm0.04$ for the cluster galaxy sample, 
and $0.46\pm0.02$ for the field galaxy sample. 
These results suggest that compared to cluster galaxies, 
compact group galaxies are more dominated by MIR blue galaxies:
the fraction of MIR blue galaxies is $32.6\%\pm1.2\%$ 
in the compact group galaxy sample and $23.7\%\pm1.1\%$ in the cluster 
galaxy sample. 
This suggests that the mean stellar ages of compact group galaxies are on 
average older than those of cluster galaxies.

\begin{figure}
\centering
\includegraphics[scale=0.7]{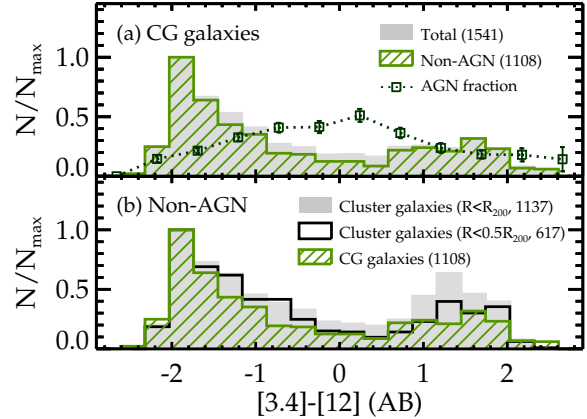}
\caption{(a) $[3.4]-[12]$ color distribution for the all 1541 (filled) and the 1108 non-AGN 
(hatched) compact group galaxies. Squares represent the fraction of 
AGN as a function of $[3.4]-[12]$. The error bars indicate the 1$\sigma$ 
deviation from 1000-times bootstrap resamplings.
(b) Comparison of the color distributions between the 1108 non-AGN compact group
galaxies (hatched), the 1137 non-AGN cluster galaxies at $R<\rvir$ (filled), 
and the 617 non-AGN cluster galaxies at $R<0.5\rvir$ (open).
}
\label{w13abhist_agn}
\end{figure}

\begin{figure*}
\centering
\includegraphics[scale=0.7]{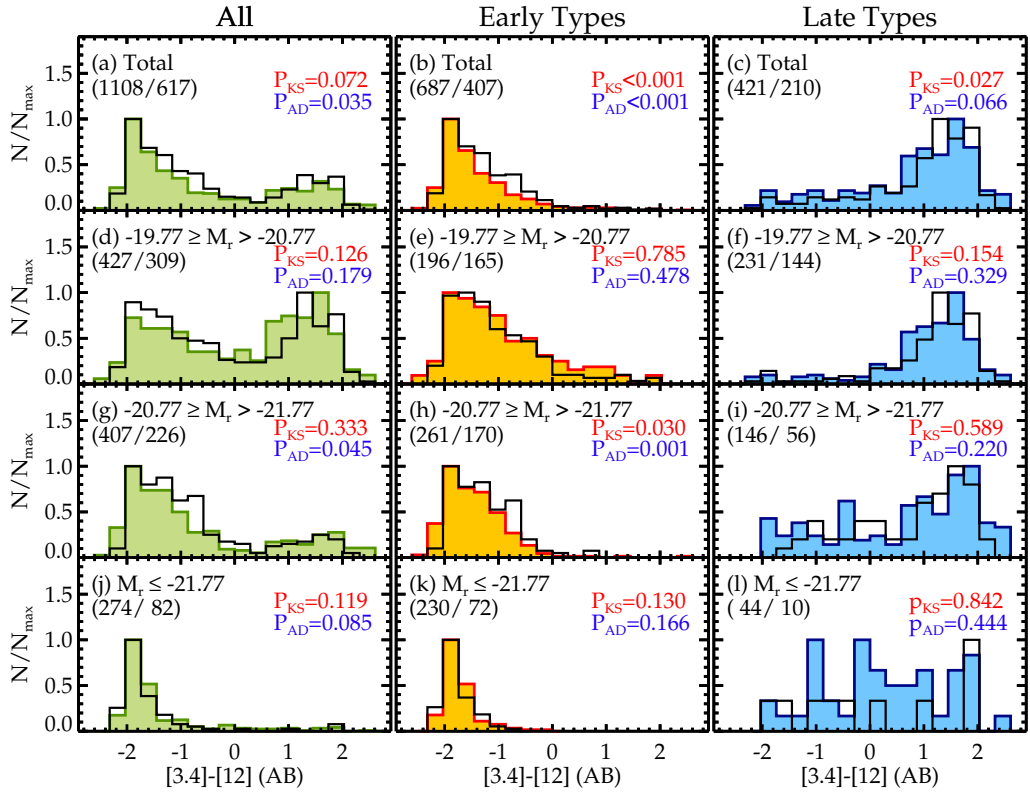}
\caption{Comparison of $[3.4]-[12]$ color distributions for 
compact group galaxies (filled histograms) with those of cluster galaxies 
(open histograms). 
The middle column is for early-type galaxies, and the right-hand column 
is for late-type galaxies. The left-hand column is the sum. 
The second row is for faint galaxies with $-19.77\geq M_r > -20.77$, 
the third row is for galaxies with $-20.77\geq M_r > -21.77$, 
and the bottom row is for bright galaxies with $M_r \leq -21.77$.  
We list $p$-values from the KS and AD k-sample tests for the two distributions 
(for compact group galaxies and for cluster galaxies) in each panel.
}
\label{w13abhist_morph_maglim}
\end{figure*}

On the other hand, because compact groups favor strong galaxy interactions, 
the presence of active galactic nuclei (AGN) should be expected
\citep{coziol+98a,coziol+98b,martinez+10,tzanavaris+14}.
Because the contribution of AGN to the MIR fluxes 
of the host galaxies is not negligible, we identify the AGN in our sample 
and remove them.
First, we identify optically selected AGN using the scheme 
given by \citet{kewley+06} who classified star-forming galaxies from AGN 
(Seyferts and LINERs) and composite galaxies in the [NII]/H$\alpha$ versus [OII]/H$\beta$ diagram (see their Figure 1). We remove 281 AGN and 149 composite 
galaxies from the sample of 1541 compact group galaxies. 
Second, we identify MIR selected AGN using the classification method 
introduced by \citet{mateos+12} who defined the AGN wedge in the \textit{WISE} 
$[3.4]-[4.6]$ versus $[4.6]-[12]$ color-color diagram (see their Figure 2). 
We find 8 MIR selected AGN in the compact group galaxy sample. 
Among the 8 MIR selected AGN, 5 are also optically selected AGN. 
In total, 433 galaxies (28\%) are classified as AGN-host galaxies among the 1541 
compact group galaxies.

\begin{figure*}
\centering
\includegraphics[scale=0.7]{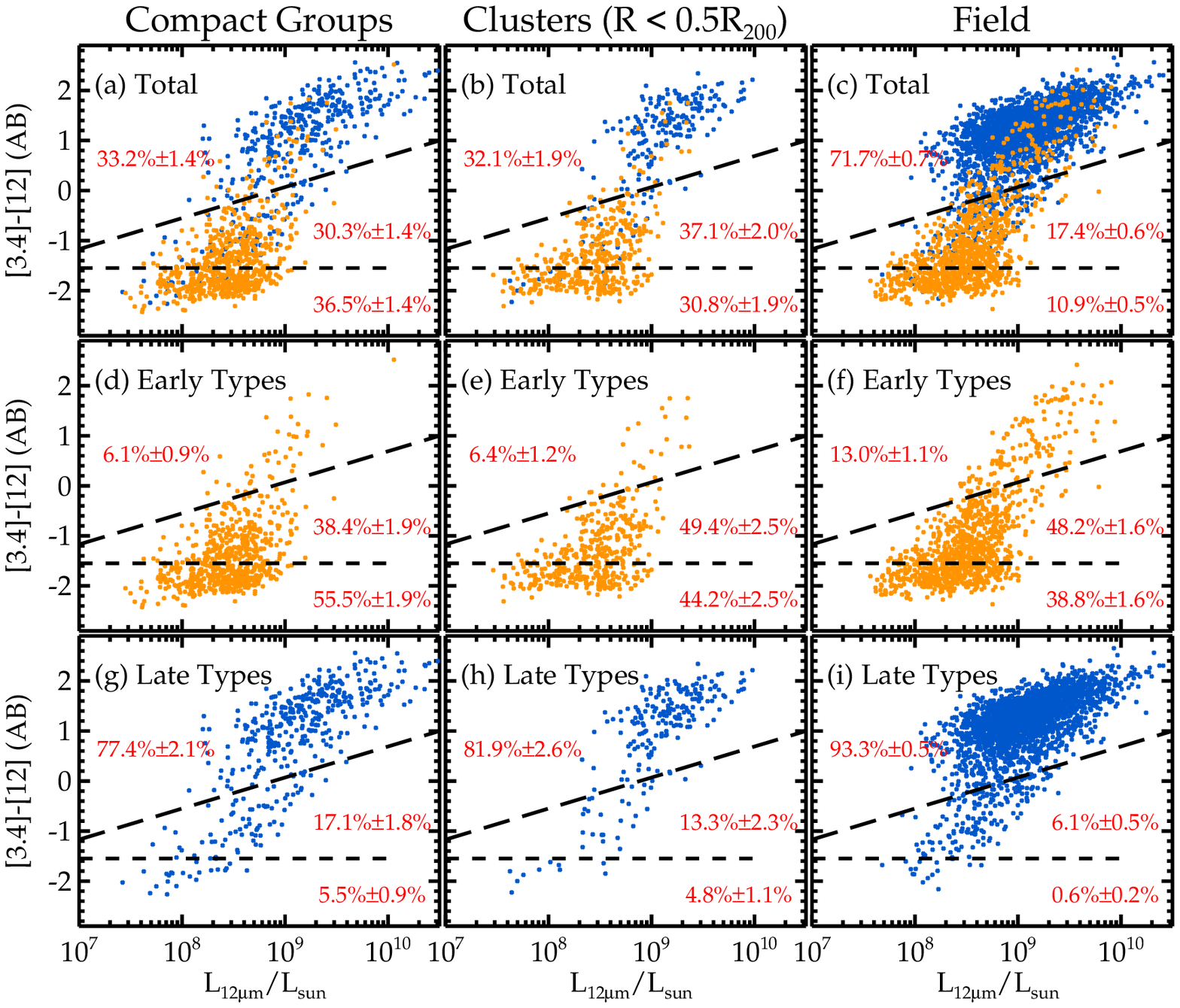}
\caption{MIR color vs. luminosity diagrams for compact group 
galaxies (left), cluster ($R<0.5\rvir$) galaxies (middle), 
and field galaxies (right). 
Orange and blue dots represent early- and late-type galaxies, respectively. 
We divide the galaxies into three classes following the classification scheme 
of \citet{lee+15}: MIR star-forming 
sequence galaxies (above the long-dashed lines), MIR blue cloud galaxies
(below short-dashed lines), and MIR green valley galaxies (between the two lines). 
Middle and bottom rows are for early- and late-type galaxies in 
different environments. We list the fractions of different MIR classes in each panel.
}
\label{mircld_env}
\end{figure*}


\begin{deluxetable*}{ccrcrccc}
\tablecolumns{8}
\tablewidth{0pc}
\tablecaption{\textit{WISE}-related Physical Parameters of Compact Group Galaxies}
\tablehead{
\colhead{Host Group} & \colhead{Galaxy ID} & \colhead{$[3.4]-[12]^a$} & 
\colhead{(S/N)$_{12}^b$} & 
\colhead{${\rm log}(L_{12}/L_{\odot})^c$} & 
\colhead{AGN$^d$} & \colhead{$M_{3.4}^e$} & 
\colhead{log($M_{\rm star}/M_{\rm sun})^f$}
}
\startdata
V1CG001 & 1237648705657307198  & $-9.99\pm0.00$ &  -9.9   & $-9.99\pm0.00$ &  0 &  $-9.99\pm0.00$ &  $0.00_{-0.00}^{+0.00}$ \\
V1CG001 & 1237648705657307347  & $0.00\pm0.00$ &   1.4    & $8.60\pm0.00$ &  0  & $-23.04\pm0.03$ &  $9.92_{-0.05}^{+0.07}$ \\
V1CG001 & 1237648705657307315  & $-0.92\pm0.00$ &   0.2    & $8.50\pm0.00$ &  0  & $-23.71\pm0.04$ &  $10.61_{-0.17}^{+0.12}$ \\
V1CG002 & 1237661126155436166  & $-0.87\pm0.05$ &  26.5   & $8.40\pm0.02$ &  1  & $-23.42\pm0.02$ &  $10.70_{-0.15}^{+0.12}$ \\
V1CG002 & 1237661126155436169  & $0.69\pm0.03$ &  44.9   & $8.95\pm0.01$ &  1  & $-23.22\pm0.02$ &  $10.63_{-0.10}^{+0.08}$ \\
V1CG002 & 1237661126155436164  & $-2.03\pm0.05$ &  23.2   & $8.61\pm0.02$ &  0  & $-25.10\pm0.02$ &  $11.59_{-0.13}^{+0.07}$ \\
V1CG003 & 1237661136886431890  & $0.81\pm0.11$ &   9.3    & $8.71\pm0.05$ &  0  & $-22.51\pm0.03$ &  $10.13_{-0.12}^{+0.09}$ \\
V1CG003 & 1237661136886497282  & $2.04\pm0.05$ &  31.7   & $9.67\pm0.01$ &  1  & $-23.67\pm0.04$ &  $10.20_{-0.13}^{+0.09}$ \\
V1CG003 & 1237661136886497353  & $0.41\pm0.10$ & 10.6    & $8.74\pm0.04$ &  0  & $-23.00\pm0.03$ &  $10.14_{-0.08}^{+0.08}$ 
\enddata
\label{tab2}
\tablecomments{The full table is available in the online journal. 
A portion is shown here for guidance regarding its form and content.
$^a$ The \textit{ALLWISE} profile-fitting AB magnitudes. $-9.99$ indicates 
no \textit{ALLWISE} photometry.  \\
$^b$ The signal-to-ratio at 12 $\micron$. $-9.9$ indicates no \textit{ALLWISE} photometry. When (S/N)$_{12}\leq3$, we assign zero errors to 
$[3.4]-[12]$ and ${\rm log}(L_{12}/L_{\odot})$.\\
$^c$ 12 $\micron$ luminosities. $-9.99$ indicates no \textit{ALLWISE} photometry. \\
$^d$ AGN classification (0: non-AGN, 1: optical AGN, 2: MIR AGN, and 3: optical$+$MIR AGN) \\
$^e$ 3.4 $\micron$ absolute magnitudes. $-9.99\pm0.000$ indicates no \textit{ALLWISE} photometry. \\
$^f$ Stellar masses calculated using $H_0=70~\kms~{\rm Mpc^{-1}}$. 0.00 means no measurement. 
}
\end{deluxetable*}

We compare the MIR color distribution of all 1541 compact group galaxies 
with that of 1108 non-AGN compact group galaxies in Figure \ref{w13abhist_agn}(a). 
The difference between the two distributions is remarkable at 
$-1.5\lesssim[3.4]-[12]\lesssim0.5$ where the AGN fraction is larger than 30\%.
This color range is similar to the \textit{WISE} infrared transition zone, 
$-1.71\lesssim[3.4]-[12]\lesssim0.02$, defined by \citet{alatalo+14}\footnote{
This color range is converted from the $0.8<[4.6]-[12]<2.4$ in Vega magnitude 
in \citet{alatalo+14}.}.
They showed that Seyferts and LINERs are representatives of the zone, 
which is consistent with our result. 

In Figure \ref{w13abhist_agn}(b), we compare the MIR color distribution 
of non-AGN compact group galaxies with that of non-AGN cluster galaxies. 
In this comparison, we use 1137 non-AGN cluster galaxies ($R<\rvir$) after removing 
324 AGN-host galaxies (22\%) from the 1461 cluster galaxy sample.
To statistically examine the difference in the color distributions of the two samples, 
we compute $p$-values from the Kolmogorov-Smirnov (KS) test ($P_{\rm KS}$) 
and the Anderson-Darling (AD) k-sample test ($P_{\rm AD}$). 
The $P_{\rm KS}$ and $P_{\rm AD}$ indicate the 
probability of rejecting the null hypothesis that the two samples are drawn from 
the same parent sample. The two tests on the two color distributions give 
the $P_{\rm KS}$ and $P_{\rm AD}$ of $\ll0.01$, rejecting the null hypothesis at a significance of $>3\sigma$ level.

In Figure \ref{w13abhist_agn}(b), we use another comparison sample of 
617 cluster galaxies located at $R<0.5\rvir$. 
In comparison with the $R<\rvir$ cluster galaxy sample, the $R<0.5\rvir$ cluster 
galaxy sample displays the MIR color distribution more colsely to that of 
the compact group galaxy sample ($P_{\rm KS}=0.076$ and $P_{\rm AD}=0.035$; 
the null hypothesis is rejected at the $<1.8\sigma$ level).
The fractions of MIR red (i.e., $[3.4]-[12]\geq-0.3$) galaxies in the compact group
galaxy sample is $35.4\%\pm1.4\%$. This is similar to the $34.7\%\pm1.8\%$ 
in the $R<0.5\rvir$ cluster galaxy sample, but smaller than the $43.9\%\pm1.5\%$ 
in the $R<\rvir$ cluster galaxy sample. The mean $[3.4]-[12]$ for the compact 
group galaxy sample ($-0.57\pm0.04$) is similar to the $-0.52\pm0.06$ for 
the $R<0.5\rvir$ cluster galaxy sample, but smaller than the $-0.30\pm0.04$ for the 
$R<\rvir$ cluster galaxy sample. 
The early-type galaxy fraction in the $R<0.5\rvir$ cluster galaxy sample 
($66.0\%\pm1.9\%$) is similar to that in the compact group galaxy sample 
($62.0\%\pm1.5\%$), while the fraction is $57.5\%\pm1.5\%$ in the $R<\rvir$ cluster 
galaxy sample.
These results suggest that compact groups are 
composed of galaxy populations similar to those in cluster central regions.

However, we find that the fraction of MIR green (i.e., $-1.55\leq[3.4]-[12]<-0.3$) 
galaxies is smaller in the compact group galaxy sample ($28.5\%\pm1.3\%$) than 
in the $R<0.5\rvir$ cluster galaxy sample ($34.5\%\pm1.9\%$). 
This may suggest that galaxy transition occurs faster in compact groups than 
in cluster environments. 
Hereafter, we use only the $R<0.5\rvir$ cluster galaxy sample in the comparison 
with the compact group galaxy sample.

Figure \ref{w13abhist_morph_maglim}(a) shows again the MIR color distributions
of the 1108 non-AGN compact group galaxies and of the 617 non-AGN cluster galaxies. 
In Figure \ref{w13abhist_morph_maglim} we divide these two galaxy samples 
into early-type and late-type galaxy samples.
In Figure \ref{w13abhist_morph_maglim}(b) we compare 
the MIR color distribution of the 687 compact group early-type 
galaxies with that of the 407 cluster early-type galaxies.
The $P_{\rm KS}$ and $P_{\rm AD}$ values of 
$<0.001$ reject the null hypothesis for the two color distributions 
at a $>3\sigma$ significance. 
We find that the fraction of MIR green galaxies is smaller in the compact group
early-type galaxy sample ($37.6\%\pm1.8\%$) than in the cluster early-type galaxy 
sample ($47.4\%\pm2.5\%$), and that the mean $[3.4]-[12]$ value is smaller 
in the compact group early-type galaxy sample ($-1.40\pm0.03$) than in 
the cluster early-type galaxy sample ($-1.28\pm0.04$).

In Figure \ref{w13abhist_morph_maglim}(c) we compare the color distribution of 
the 421 compact group late-type galaxy sample with that of the 210 cluster late-type galaxy 
sample. The difference in the color distribution for the two late-type galaxy samples 
is not highly significant ($<1.9\sigma$), compared to the case for early-type galaxy samples. However, the mean $[3.4]-[12]$ value is smaller in the compact group
late-type galaxy sample ($0.80\pm0.06$) than in the cluster late-type galaxy sample
($0.95\pm0.07$). The MIR green galaxy fraction is larger in the compact group
late-type galaxy sample ($13.8\%\pm1.7\%$) than in the cluster late-type galaxy 
sample ($9.5\%\pm1.9\%$).

Because galaxy properties are dependent on their stellar masses (or luminosities),
we compare the MIR color distributions of compact group galaxies and 
cluster galaxies by using a fixed $M_r$. We use $M_r$ instead of stellar masses 
because the former is available for all galaxies in both compact group and cluster
galaxy samples.
We find that the null hypothesis is rejected in the comparison 
for early-type galaxies with $-20.77\geq M_r > -21.77$ at the $\gtrsim2.2\sigma$ level 
(panel h; $P_{\rm KS}=0.03$ and $P_{\rm AD}=0.001$). 
For the total galaxy sample with $-20.77\geq M_r > -21.77$ (panel g), 
$P_{\rm AD}$ rejects the null hypothesis at the $2\sigma$ level, 
but $P_{\rm KS}$ does not.
The other cases show lower significance of rejection. 
This indicates that the difference in the MIR color distribution between 
the compact group galaxies and the cluster galaxies is mainly attributed 
to the galaxies with $-20.77\geq M_r > -21.77$.
We examine galaxy properties using the luminosity-limited subsamples 
in the following analysis, but find no significant difference from the results 
based on the total sample. 
We therefore present the results in the following analysis based on the total sample 
for better statistics.


\begin{deluxetable*}{cccccccc}
\tablecolumns{8}
\tablewidth{0pc}
\tablecaption{Environmental Parameters of Compact Groups}
\tablehead{
\colhead{Group ID$^a$} & \colhead{R.A.$^a$} & \colhead{Decl.$^a$} & 
\colhead{Redshift$^a$} & \colhead{$N_{\rm mem}^b$} & \colhead{$\nbg^c$} & 
\colhead{$\efrac^d$} & 
\colhead{$\sigma_{\rm CG}~(\kms)^e$}
}
\startdata
 V1CG001 & 198.227173  &  1.012775   & $0.0723\pm0.0008$ & 3 & -99 & 0.67 & $482\pm  122$ \\
 V1CG002 & 139.935760  & 33.744308  & $0.0202\pm0.0012$ & 3 & 11  & 0.67  & $866\pm  228$ \\
 V1CG003 & 154.741531  & 37.298065  & $0.0480\pm0.0003$ & 3 &   5  & 0.00  & $210\pm   37$ \\
 V1CG004 & 158.222275  & 12.086633  & $0.0330\pm0.0004$ & 3 & 11  & 0.67  & $242\pm   69$ \\
 V1CG005 & 127.709404  & 28.573534  & $0.0657\pm0.0001$ & 3 &   1  & 0.33  & $26\pm    4$ \\
 V1CG006 & 149.408493  & 37.324886  & $0.0611\pm0.0002$  & 3 &  4  & 0.33  & $100\pm   26$ \\
 V1CG007 & 240.816635  & 45.356598  & $0.0420\pm0.0002$  & 3 &  3  & 0.33  & $89\pm   15$ \\
 V1CG008 & 239.193481  & 46.345490  & $0.0420\pm0.0009$  & 3 &  3  & 1.00  & $483\pm   83$ \\
 V1CG009 & 204.321213  & 45.249256  & $0.0620\pm0.0003$  & 3 &  7  & 0.67  & $189\pm   47$ \\
 V1CG010 & 161.923065  & 38.933136  & $0.0360\pm0.0004$  & 4 & 14 &  0.50  & $272\pm   75$
\enddata
\label{tab3}
\tablecomments{The full table is available in the online journal. 
A portion is shown here for guidance regarding its form and content.\\
$^a$ Group ID from Table 5 of \citet{sohn+16}. \\
$^b$ The number of member galaxies in each compact group. \\ 
$^c$ The number of neighboring galaxies ($M_r < -19.77$) in the comoving cylinder. 
For 59 compact groups near the lower and upper redshift limits, $\nbg$ are 
not assigned ($\nbg=-99$).  \\
$^d$ The fraction of early-type galaxies among member galaxies of a compact group.\\
$^e$ The rest-frame line-of-sight velocity dispersion of galaxies in each compact group
and the 1$\sigma$ error derived from 1000-times bootstrap resamplings,  
from Table 5 of \citet{sohn+16}. 
}
\end{deluxetable*}

\subsection{MIR Color-Luminosity Diagram}

In Figure \ref{mircld_env} we investigate the $[3.4]-[12]$ versus 
12 $\micron$ luminosity distribution of galaxies in compact groups, clusters, and field. 
In the $[3.4]-[12]$ versus 12 $\micron$ luminosity diagram, galaxies are divided 
into three classes following the classification scheme of \citet{lee+15}: 
MIR star-forming sequence galaxies (above the inclined dashed lines\footnote{
$[3.4]-[12] = {\rm log}(L_{12}/L_{\rm sun})\times0.62-5.51$. See Section 5.1 of 
\citet{lee+15} for details.}), 
MIR blue cloud galaxies ($[3.4]-[12]<-1.55$), and MIR green valley galaxies 
located between the two classes.
In this section we use this classification scheme to calculate the fractions of 
different MIR classes.
In particular, MIR green valley galaxies are not necessarily 
``MIR green" in the sense of Section 3.1.

In the three panels in the top row of Figure \ref{mircld_env}, 
we plot 1108 compact group galaxies, 617 cluster galaxies, and 3854 field galaxies,
separately. All these galaxies are not AGN-host galaxies. 
We find that the fraction of MIR blue cloud galaxies is larger in the compact group galaxy 
sample ($36.5\%\pm1.4\%$) than in the cluster galaxy sample ($30.8\%\pm1.9\%$) 
and in the field galaxy sample ($10.9\%\pm0.5\%$). 
In the cluster galaxy sample and the field galaxy sample, the fractions of 
MIR blue cloud galaxies are smaller than the fractions of MIR green valley galaxies.
In contrast, in the compact group galaxy sample, the fraction of MIR blue cloud 
galaxies is larger than the fraction of MIR green valley galaxies ($30.3\%\pm1.4\%$).

When we consider only early-type galaxies, as shown in the middle row of 
Figure \ref{mircld_env}, we obtain similar results. 
The fraction of early-type galaxies is 62.0\% (687/1108) in the compact group
galaxy sample, 66.0\% (407/617) in the cluster galaxy sample, and 
26.5\% (1023/3854) in the field galaxy sample.
Among the three early-type galaxy samples, the compact group early-type galaxy 
sample shows the highest MIR blue cloud galaxy fraction.
In addition, 
in the cluster early-type galaxy sample and in the field early-type galaxy sample, 
the fractions of MIR blue cloud galaxies is smaller than those of MIR green valley 
galaxies. However, in the compact group early-type galaxy sample, the fraction 
of MIR blue cloud galaxies ($55.5\%\pm1.9\%$) is significantly larger than 
the fraction of MIR green valley galaxies ($38.4\%\pm1.9\%$). 
These results suggest not only that early-type galaxies in compact groups are 
on average older than those in other environments, but also that 
the timescales for early-type galaxies to stay in the MIR green valley are 
shorter in compact groups than in other environments.

In the bottom row of Figure \ref{mircld_env}, we plot only late-type galaxies.
The fraction of MIR star-forming sequence galaxies decreases from 
the field ($93.3\%\pm0.5\%$) to clusters ($81.9\%\pm2.6\%$) 
to compact groups ($77.4\%\pm2.1\%$). 
Thus, the fractions of MIR blue cloud galaxies and of MIR green valley 
galaxies are smallest in the field ($0.6\%\pm0.2\%$ and $6.1\%\pm0.5\%$) and 
largest in compact groups ($5.5\%\pm0.9\%$ and $17.1\%\pm1.8\%$). 
The smallest fraction of MIR star-forming sequence galaxies 
in compact groups is consistent with the case that the star formation quenching for 
late-type galaxies occurs more effectively in compact groups than in clusters 
and in the field.
We note that the results in this section do not change significantly 
if we use $-1.55\leq[3.4]-[12]<-0.3$ as the criterion to select MIR green galaxies.
In Table \ref{tab2}, we list the \textit{WISE}-related parameters and AGN 
classes of compact group galaxies.

\subsection{Environments of Compact Group Galaxies}\label{env_cg}

To investigate the environmental effects on galaxy properties in compact groups, 
we use three kinds of environmental parameters. 
The first parameter is the number of neighboring galaxies ($\nbg$) of compact groups.
\citet{sohn+16} calculated $\nbg$ around compact groups 
within a comoving cylinder defined by projected group centric radius $R = 1$ Mpc 
and $\Delta V = | V_{\rm gal} - V_{\rm gr} | = 1000$ km s$^{-1}$, 
where $V_{\rm gal}$ and $V_{\rm gr}$ are a radial velocity of a galaxy and 
a mean velocity of member galaxies in a compact group.
In the $\nbg$ count, \citet{sohn+16} used galaxies in the V2 sample ($M_r<-20.77$ 
and $0.01<z<0.115$). In this study, we recalculate $\nbg$ for 670 compact groups
using galaxies in the V1 sample ($M_r<-19.77$ and $0.01<z<0.0741$). 
We exclude the compact group members in the $\nbg$ count. 
We do not assign $\nbg$ to 59 compact groups close to the lower and 
upper redshift limits because their $\nbg$ can be underestimated, 
resulting in that we assign $\nbg$ to 611 compact groups.

The second environmental parameter is the fraction of early-type member 
galaxies in compact groups ($\efrac$).
\citet{bitsakis+10,bitsakis+11,bitsakis+15} used this $\efrac$ 
to divide compact groups into dynamically old and young systems
based on the assumption that dynamically old systems are dominated 
by early-type galaxies that could form through repeated interactions between 
member galaxies.
Thus, $\efrac$ is the parameter reflecting the internal environment of compact groups. 

The third environmental parameter is the rest-frame velocity dispersions of 
compact group member galaxies ($\sigma_{\rm CG}$). 
We adopt the $\sigma_{\rm CG}$ from Table 5 in \citet{sohn+16}.
\citet{sohn+16} found that compact groups with low $\sigma_{\rm CG}$ 
($\lesssim100$ km s$^{-1}$) show features of ongoing interactions among member 
galaxies, but that compact groups with $\sigma_{\rm CG}\gtrsim300$ km s$^{-1}$ 
do not. 
Thus, $\sigma_{\rm CG}$ is another parameter that reflects the internal environment 
of compact groups. 

\begin{figure}[b]
\centering
\includegraphics[scale=0.7]{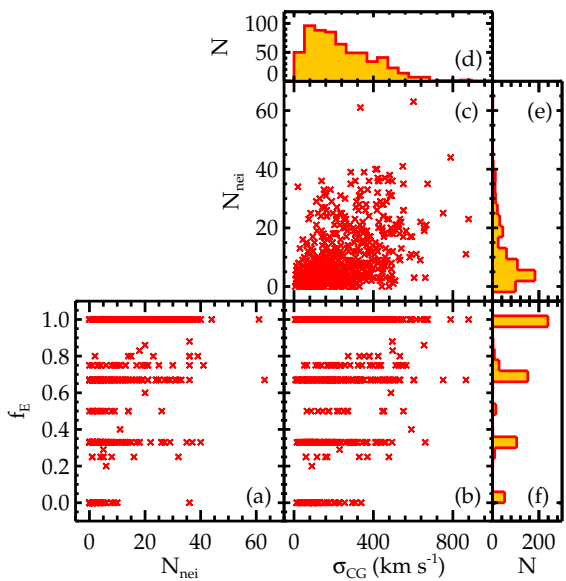}
\caption{Relation between the three environmental parameters for compact groups: 
(a) $\efrac$ vs. $\nbg$, (b) $\efrac$ vs. $\sigma_{\rm CG}$, and 
(c) $\nbg$ vs. $\sigma_{\rm CG}$ diagrams. (d$-$f) Histograms for 
the three parameters.}
\label{cg_nc_efrac}
\end{figure}

We note that we use all the member galaxies regardless of 
12 $\micron$ detection
when we calculate $f_{\rm E}$ and $\sigma_{\rm CG}$.
In Table \ref{tab3}, we list the ID, R.A., declination, redshift, 
as well as three environmental parameters of the 670 compact groups.
Figure \ref{cg_nc_efrac} shows the relations between the three environmental 
parameters. We plot only 611 systems with complete $\nbg$.
The $\nbg$ histogram peaks at $\nbg\simeq3-6$, 
and stretches out to more than $\nbg=40$.
The $\efrac$ mainly has discontinuous values (e.g., 0, 1/3, 2/3, and 1)
because the majority (495/611, 81\%) 
of our systems have three member galaxies.
The $\sigma_{\rm CG}$ histogram peaks at $\sim80$ km s$^{-1}$, and 
stretches out to more than 600 km s$^{-1}$. The median $\sigma_{\rm CG}$ 
is $\sim200$ km s$^{-1}$. 

\begin{figure*}
\centering
\includegraphics[scale=0.7]{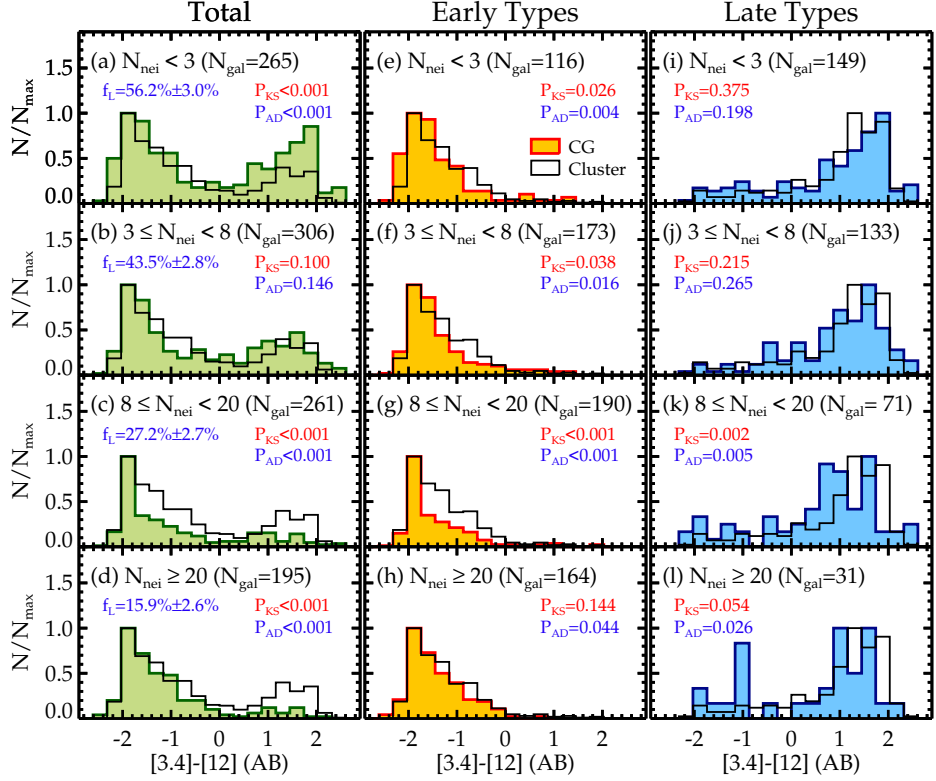}
\caption{Dependence of the $[3.4]-[12]$ color of compact group galaxies 
on the number of neighboring galaxies ($\nbg$). In the left-hand column, 
we list the fraction of late-type galaxies ($f_{\rm L}$).  
In the middle and right-hand columns, we plot early-type galaxies and late-type galaxies,
separately. For comparison,
we also plot the color distributions for cluster galaxies (open histograms).
}
\label{w13abhist_nbg}
\end{figure*}

\subsubsection{Environmental Dependence on $\nbg$}

To examine the effects of $\nbg$ on galaxy properties in compact groups,
we investigate how the MIR color distribution of compact group galaxies 
varies with $\nbg$. 
Figure \ref{w13abhist_nbg} shows the MIR color histograms of the galaxies  
in different $\nbg$ bins. 
The left-hand column shows that 
the fraction of MIR red (i.e., $[3.4]-[12]>-0.3$) 
galaxies dramatically decreases as $\nbg$ increases. 
The MIR red galaxies are mainly late-type galaxies.
The fraction of late-type galaxies gradually decreases from $56.2\%\pm3.0\%$ 
(panel a) to $15.9\%\pm2.6\%$ (panel d) as $\nbg$ increases. 
For comparison, we also plot the MIR color distribution of cluster galaxies 
as open histograms. 
We find that the fraction of late-type galaxies in the $\nbg\geq8$ compact groups 
is smaller than the fraction of late-type galaxies ($34.0\%\pm1.9\%$) in the cluster 
galaxy sample. 

We investigate the MIR color distributions for early- and late-type galaxies
separately in the middle and right columns in Figure \ref{w13abhist_nbg}.
The KS and AD k-sample tests for early-type galaxy samples in the four different
$\nbg$ bins (e$-$h) reject the null hypothesis at the $<1.3\sigma$ level.
They also reject the null hypothesis for late-type galaxy samples in the four different 
$\nbg$ bins (i$-$l) at the $<1.8\sigma$ level. 
These imply that $\nbg$ does not affect the MIR colors of early- and late-type 
galaxies in compact groups directly.
Thus, the $\nbg$-dependence of MIR colors shown in the left column (a$-$d) 
results from the fact that compact groups located in denser regions 
(larger $\nbg$) tend to have more early-type member galaxies with MIR blue colors.

The middle column in Figure \ref{w13abhist_nbg} shows that  
compact group early-type galaxies have MIR colors bluer than 
cluster early-type galaxies regardless of $\nbg$.
The color difference is significant ($\gtrsim2.1\sigma$) in the $\nbg<20$ compact groups, 
but it is marginal ($\sim1.5\sigma$) in the $\nbg\geq20$ groups. 
We find no color difference between cluster late-type galaxies 
and compact group late-type galaxies at the $\nbg<8$ groups.
However, late-type galaxies at the $\nbg\geq8$ groups have MIR colors bluer 
than cluster late-type galaxies.
The significance of the color difference is $\sim$2$-$3$\sigma$.

\subsubsection{Environmental Dependence on $\efrac$}

\begin{figure}[h]
\centering
\includegraphics[scale=0.7]{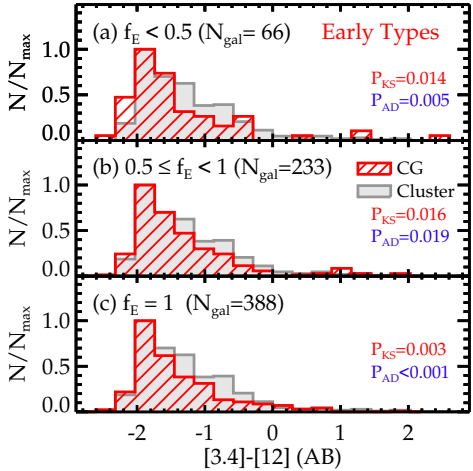}
\caption{
MIR color distributions for early-type galaxies in the (a) $\efrac<0.5$, 
(b) $0.5\leq\efrac<1$, and (c) $\efrac=1$ compact groups.
For comparison, we also plot the (filled) histograms for the early-type galaxies 
in our cluster sample. 
}
\label{w13_efrac_egal}
\end{figure}

\begin{figure}[ht]
\centering
\includegraphics[scale=0.7]{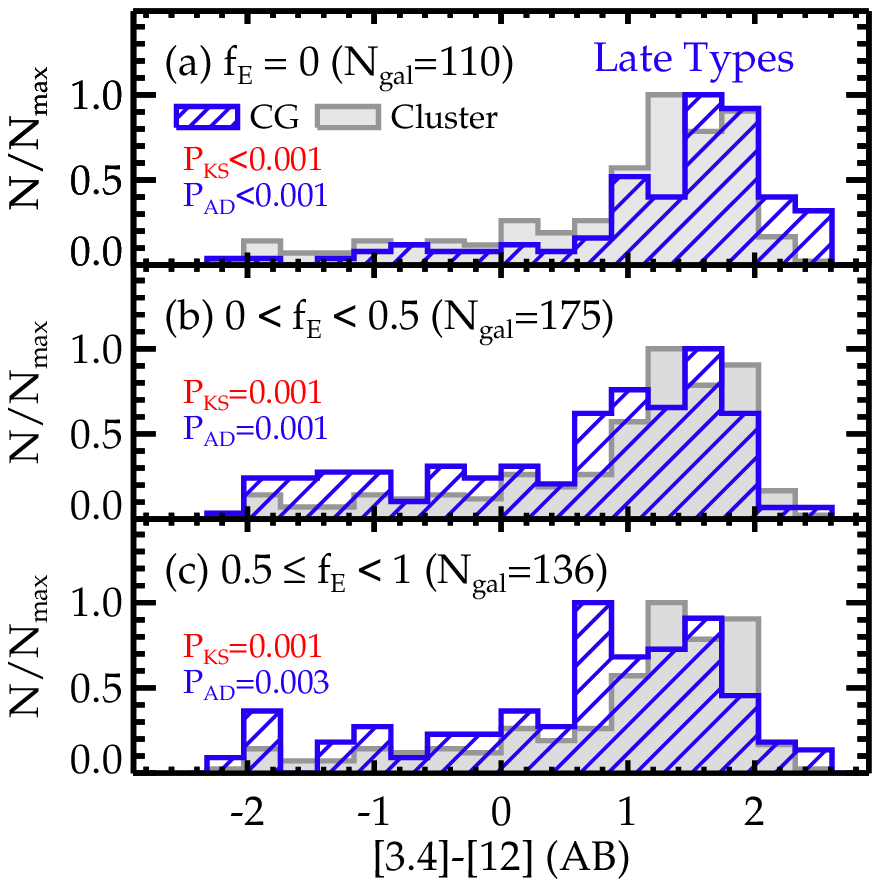}
\caption{
MIR color distributions for late-type galaxies in the (a) $\efrac=0$,
(b) $0<\efrac<0.5$, and (c) $0.5\leq\efrac<1$ compact groups. 
For comparison, we also plot the (filled) histograms for the late-type galaxies 
in our cluster sample.}
\label{w13_efrac_lgal}
\end{figure}

\begin{figure*}
\centering
\includegraphics[scale=0.7]{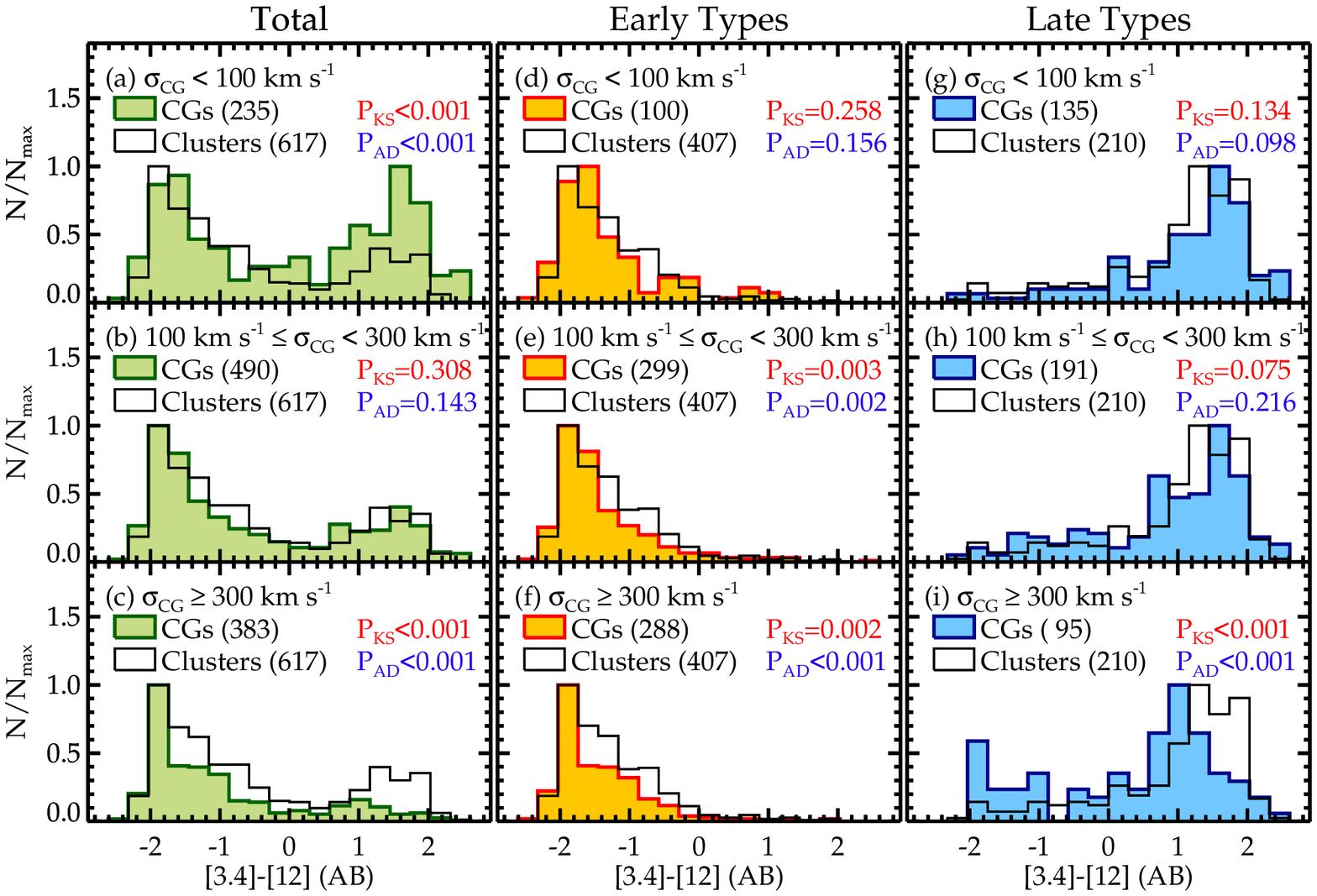}
\caption{
Left: MIR color distributions for compact group galaxies (filled histograms) 
in the (a) $\sigma_{\rm CG}<100$ km s$^{-1}$, 
(b) 100 km s$^{-1} \leq \sigma_{\rm CG} <$ 300 km s$^{-1}$, 
and (c) $\sigma_{\rm CG}\geq300$ km s$^{-1}$ compact groups.
Open histograms represent cluster galaxies. 
We plot early-type galaxies in the middle column (d$-$f) and 
late-type galaxies in the right column (g$-$i).
}
\label{w13abhist_vdispcg}
\end{figure*}

Figure \ref{w13_efrac_egal} shows the MIR color distributions 
of compact group early-type galaxies in different $\efrac$ bins: $\efrac<0.5$, 
$0.5\leq\efrac<1$, and $\efrac=1$. 
We find no significant differences in the MIR colors between the three $\efrac$ bins.
The mean $[3.4]-[12]$ value is $-1.49\pm0.08$ at $\efrac<0.5$, 
$-1.44\pm0.04$ at $0.5\leq\efrac<0.1$, and $-1.43\pm0.03$ at $\efrac=1$, respectively. 
These mean values agree well within $1\sigma$ errors. 
The KS and the AD k-sample tests do not reject the null hypothesis, either.

\citet{bitsakis+11} showed that early-type galaxies in dynamically young
($\efrac<0.25$) compact groups have bluer NUV$-r$ colors than
those in dynamically old ($\efrac>0.25$) groups. 
However, the number of early-type galaxies 
in their dynamically young groups is only three. 
In our sample, there are no early-type galaxies belonging to the $\efrac<0.25$ groups, 
and only six early-type galaxies belonging to the $\efrac<0.3$ groups. 
However, we find that the MIR color distribution for the six galaxies is 
not different significantly from that for early-type galaxies in the $\efrac<0.5$ groups.

On the other hand, we find that in all three $\efrac$ bins,
compact group early-type galaxies have MIR colors bluer than 
cluster early-type galaxies.
The $P_{\rm KS}$ and $P_{\rm AD}$ are $<0.019$, suggesting that 
the color difference between compact group and cluster early-type galaxy samples
is significant ($>2.4\sigma$).
This is consistent with the result in Figure \ref{w13abhist_nbg}.

\begin{figure*}[h]
\centering
\includegraphics[scale=0.7]{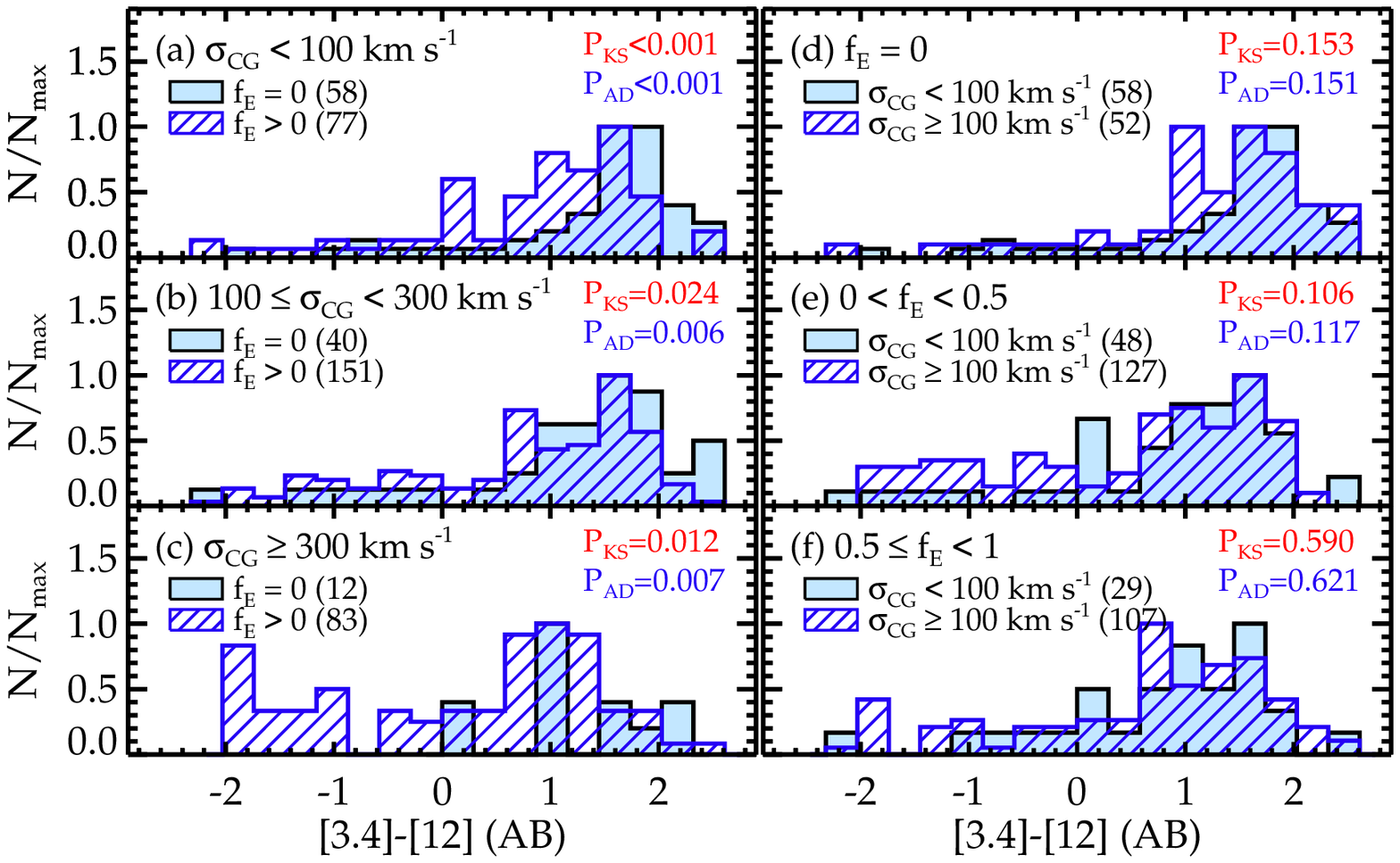}
\caption{
Left: MIR color distributions for late-type galaxies 
in the (a) $\sigma_{\rm CG}<100$ km s$^{-1}$, 
(b) $100\leq\sigma_{\rm CG}<300$ km s$^{-1}$, 
and (c) $\sigma_{\rm CG}\geq300$ km s$^{-1}$ compact groups. 
Filled and hatched histograms represent late-type galaxies in the $\efrac=0$ groups 
and those in the $\efrac>0$ groups, respectively. 
Right: The color distributions for late-type galaxies in the (d) $\efrac=0$, 
(e) $0<\efrac<0.5$, and (f) $0.5\leq\efrac<1$ compact groups. 
Filled and hatched histograms represent late-type galaxies 
belonging to the $\sigma_{\rm CG}<100$ km s$^{-1}$ groups
and the $\sigma_{\rm CG}\geq100$ km s$^{-1}$ groups, respectively.
We list $P_{\rm KS}$ and $P_{\rm AD}$ for the two histograms in each panel.
}
\label{w13abhist_vdisp_efrac}
\end{figure*}

\begin{figure*}
\centering
\includegraphics[scale=0.7]{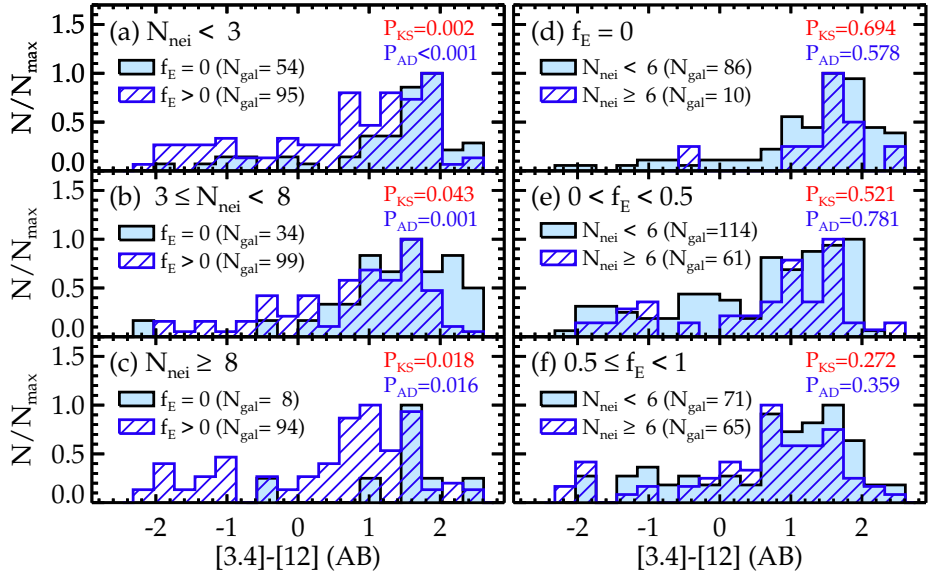}
\caption{
Left: MIR color distributions for late-type galaxies in the (a) $\nbg<3$, 
(b) $3\leq\nbg<8$, and (c) $\nbg\geq8$ compact groups. 
Filled and hatched histograms represent late-type galaxies in the $\efrac=0$ groups 
and those in the $\efrac>0$ groups, respectively. 
Right: The distributions of late-type galaxies in the (d) $\efrac=0$, 
(e) $0<\efrac<0.5$, and $0.5\leq\efrac<1$ compact groups. 
Filled and hatched histograms represent late-type galaxies 
in the $\nbg<6$ compact groups and those in the $\nbg\geq6$ compact groups,
respectively. 
}
\label{w13abhist_nbg_efrac_lgal}
\end{figure*}

\begin{figure}
\centering
\includegraphics[scale=0.7]{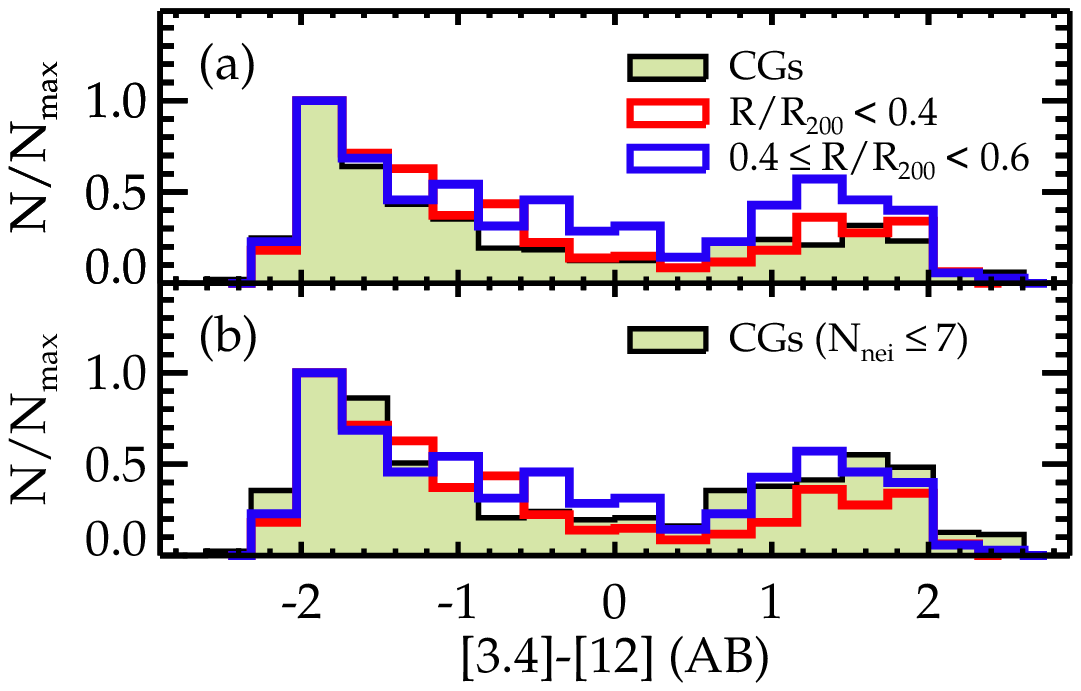}
\caption{(a) Comparison of MIR color distributions for compact group galaxies (filled 
histograms) and cluster galaxies (open histograms). 
Red and blue histograms are for cluster galaxies with $R/R_{200}<0.4$
and $0.4\leq R/R_{200} < 0.4$, respectively.
(b) We only plot galaxies belonging to $\nbg\leq7$ compact groups.
}
\label{comp_w13abhist_cluster}
\end{figure}

Figure \ref{w13_efrac_lgal} shows the MIR color distributions
of compact group late-type galaxies in different $\efrac$ bins:
$\efrac=0$, $0<\efrac<0.5$, and $0.5\leq\efrac<1$. 
For comparison, we also plot the color distributions for the cluster late-type galaxy sample.
Late-type galaxies in the $\efrac=0$ compact groups have  
MIR colors redder than cluster late-type galaxies. 
However, late-type galaxies in the $\efrac>0$ groups have MIR colors bluer 
than cluster late-type galaxies.
In all three bins, $P_{\rm KS}$ and $P_{\rm AD}$ are $\leq0.03$,
indicating that the color differences are significant ($>3\sigma$). 
On the other hand, we find that 
MIR colors of late-type galaxies in the $0<\efrac<0.5$ groups are not different from 
those in the $0.5\leq\efrac<1$ groups ($P_{\rm KS}=0.835$ and $P_{\rm AD}=0.664$). 
The mean $[3.4]-[12]$ values of late-type galaxies in 
the $0<\efrac<0.5$ groups ($0.58\pm0.09$) and in the $0.5\leq\efrac<1$ groups
($0.64\pm0.10$) are similar, but smaller than the $1.32\pm0.09$ 
in the $\efrac=0$ groups.

These results suggest that star formation activity of late-type galaxies 
is suppressed more efficiently in the $\efrac>0$ compact groups than in clusters. 
However, late-type galaxies in the $\efrac=0$ groups have comparable or higher 
star formation activity than that of cluster late-type galaxies.
It is interesting that late-type member galaxies in compact groups show 
a hint of star formation quenching only
when the compact groups have early-type member galaxies (see Section 4.3
for detailed discussion).

\subsubsection{Environmental Dependence on $\sigma_{\rm CG}$}

Figure \ref{w13abhist_vdispcg} shows how the MIR colors  
of compact group galaxies depend on $\sigma_{\rm CG}$. 
We divide compact groups into low 
( $< 100$ km s$^{-1}$), intermediate (100$-$300 km s$^{-1}$), 
and high ($\geq300$ km s$^{-1}$) $\sigma_{\rm CG}$ systems.
In the left column, we find that the fraction of MIR red (i.e., $[3.4]-[12]>-0.3$) 
galaxies decreases as $\sigma_{\rm CG}$ increases. 
This is because that the fraction of late-type galaxies is smaller in 
higher $\sigma_{\rm CG}$ systems.
This trend is similar to the $\nbg$-dependence in Figure \ref{w13abhist_nbg}.

The MIR colors of compact group early-type galaxies do not depend on 
$\sigma_{\rm CG}$.
The MIR color distribution of early-type galaxies in the $\sigma_{\rm CG}<100$
km s$^{-1}$ groups does not differ from that of cluster early-type galaxies. 
However, the MIR colors of early-type galaxies in  
the $\sigma_{\rm CG}\geq100$ km s$^{-1}$ groups are significantly ($>3\sigma$) 
different (bluer) from those of cluster early-type galaxies.
 
The MIR colors of late-type galaxies are bluer 
in the $\sigma_{\rm CG}\geq300$ km s$^{-1}$ groups 
than in the $\sigma_{\rm CG}<300$ km s$^{-1}$ groups.
The $P_{\rm KS}$ and $P_{\rm AD}$ confirm the color difference with 
a significance $\gtrsim3\sigma$ level. 
Furthermore, the MIR colors of late-type galaxies 
in the $\sigma_{\rm CG}\geq300$ km s$^{-1}$ groups 
are significantly ($>4\sigma$) bluer than those of cluster late-type galaxies.

\subsubsection{Which Environment Most Affects the MIR Properties of 
Compact Group Galaxies?}\label{env_params}

In Figures \ref{w13abhist_nbg}$-$\ref{w13abhist_vdispcg}, 
we find that the MIR colors of compact group early-type galaxies do not depend 
much on the three environmental parameters,
and that compact group late-type galaxies depend on
$\efrac$ and $\sigma_{\rm CG}$, but not on $\nbg$. 
In $\efrac>0$ or $\sigma_{\rm CG}\geq300$ km s$^{-1}$ compact groups, 
late-type member galaxies show bluer MIR color distributions.
\citet{bitsakis+11} showed that $\sigma_{\rm CG}$ tend to be larger in 
dynamically old ($\efrac>0.25$) compact groups than in dynamically young groups.
However, $\efrac$ and $\sigma_{\rm CG}$ for our compact groups do not have 
a significant correlation (Figure \ref{cg_nc_efrac}).
Spearman's rank correlation coefficient for 
$\efrac$ and $\sigma_{\rm CG}$ is 0.28 and the probability of obtaining 
the correlation by chance is $\ll0.001\%$, suggesting that the correlation between 
the two parameters is weak.

To examine which environmental parameter 
among $\efrac$, $\sigma_{\rm CG}$, and $\nbg$
most affects the MIR colors of galaxies, we first consider $\efrac$ and 
$\sigma_{\rm CG}$ simultaneously in Figure \ref{w13abhist_vdisp_efrac}.
In the left panels we compare MIR colors of late-type galaxies 
in the $\efrac=0$ compact groups with those of late-type galaxies 
in the $\efrac>0$ groups in three $\sigma_{\rm CG}$ bins. 
We find that MIR colors of late-type galaxies in the $\efrac>0$ groups 
are bluer than those of late-type galaxies in the $\efrac=0$ groups 
in all three $\sigma_{\rm CG}$ bins. 
The KS and AD k-sample tests reject the null hypothesis at 
a $\gtrsim2.3\sigma$ level in the three bins. 
On the other hand, in the right column of Figure \ref{w13abhist_vdisp_efrac}
we compare MIR colors of late-type galaxies 
in the $\sigma_{\rm CG}<100$ km s$^{-1}$ groups with 
those of late-type galaxies in the $\sigma_{\rm CG}\geq100$ 
km s$^{-1}$ groups in the three $\efrac$ bins. 
We find that there is no significant color difference between
the two $\sigma_{\rm CG}$ samples in all three $\efrac$ bins. 
These results suggest that the MIR colors of late-type galaxies in compact groups 
are more sensitive to $\efrac$ than $\sigma_{\rm CG}$. 

We conduct a similar analysis  
to examine whether the $\efrac$ effects 
on galaxy colors exist when $\nbg$ is fixed. 
Spearman's rank correlation coefficient for $\efrac$ and $\nbg$ is 
0.39, and the probability of obtaining the correlation by chance is $\ll0.001\%$, 
suggesting that there is a weak correlation between the two parameters.
In the left column of Figure \ref{w13abhist_nbg_efrac_lgal}, 
we compare MIR colors of late-type galaxies in the $\efrac=0$ groups with 
those in the $\efrac>0$ groups in three $\nbg$ bins.
In all three $\nbg$ bins, the MIR colors are bluer in the $\efrac>0$ groups than 
in the $\efrac=0$ groups. 
The AD k-sample test rejects the null hypothesis at the $\gtrsim2.5\sigma$ level, 
and the KS test also rejects the null hypothesis at the $>2\sigma$ level. 
On the other hand, the right column shows that there is no significant difference 
in MIR colors between the late-type galaxy samples in the $\nbg<6$ groups and 
those in the $\nbg\geq6$ groups. 
Figures \ref{w13abhist_vdisp_efrac} and \ref{w13abhist_nbg_efrac_lgal}
demonstrate that the MIR colors of late-type galaxies in compact groups change 
most significantly with $\efrac$ among the three environmental parameters,
suggesting that the $\efrac$ is the most important parameter in determining 
the MIR colors of late-type galaxies in compact groups.

\section{Discussion}\label{discuss}

\subsection{Fast Galaxy Evolution in Compact Groups}

In Section \ref{result}, we find that the MIR colors of early-type galaxies 
are bluer in compact groups than in clusters and the field. 
This trend also persists when we use several subsamples with $\nbg$, $\efrac$, 
and $\sigma_{\rm CG}$.
We also find that the late-type galaxies in the $\efrac>0$ 
(or $\sigma_{\rm CG}\geq300$ km s$^{-1}$) compact groups have MIR colors 
bluer than those of cluster late-type galaxies. 
These results imply that stellar populations in early-type galaxies 
are on average older in compact groups than in clusters, and that 
star formation activity of late-type galaxies is suppressed more efficiently 
in $\efrac>0$ compact groups than in clusters. 
These results suggest that galaxy evolution in compact groups is faster 
than in the central ($R<0.5R_{200}$) regions of clusters.

So far, several studies have concluded that galaxy evolution is faster in compact 
groups than in the field through various analyses \citep[e.g.,][]{proctor+04,de_la_rosa+07,bitsakis+10,bitsakis+11,bitsakis+16,tzanavaris+10,walker+10,walker+12,coenda+12,coenda+15,lenkic+16}. \citet{proctor+04} and \citet{mendes+05} 
found that the properties of compact groups (i.e., stellar ages and the early-type 
galaxy fraction) are more similar to those of cluster galaxies than 
those of field galaxies. However,
previous studies have not found evidence that galaxy evolution is faster 
in compact groups than in galaxy clusters. 

\citet{johnson+07} and \citet{walker+10,walker+12} showed that 
compact group galaxies show a strong bimodal \textit{Spitzer} IRAC 
$3.6-8$ $\micron$ color distribution with an evident gap at green colors.
This gap is not found in comparison samples of isolated galaxies, galaxy pairs, 
and the center of the Coma cluster. However, they found that 
the Coma infall region \citep[i.e., $0.4-0.6R_{200}$,][]{jenkins+07} 
shows the color distribution statistically similar to that of compact group galaxies. 
They interpreted these results as that the infall region and compact groups have 
a similarity in environment. 
To examine their results with our data, we select 495 galaxies at $R<0.4R_{200}$ 
(representing the central regions of clusters) and 231 galaxies at 
$0.4R_{200}\leq R<0.6R_{200}$ (representing the infall regions) from the cluster 
galaxy sample. We compare the MIR color distributions for the two cluster galaxy 
subsamples with that for the compact group galaxy sample in Figure 
\ref{comp_w13abhist_cluster}(a).
The color distribution for the compact group galaxy sample is more similar to 
that for the $R<0.4R_{200}$ cluster galaxy sample 
($P_{\rm KS}=0.211$ and $P_{\rm AD}=0.102$) 
than for the $0.4R_{200}\leq R<0.6R_{200}$ cluster galaxy sample 
($P_{\rm KS}=0.001$ and $P_{\rm AD}=0.002$). 
This result is not consistent with the results of Walker et al. 

\citet{walker+10} used a sample of 16 Hickson compact groups.
\citet{walker+12} used not only 21 Hickson compact groups but also 
16 Redshift Survey Compact Groups \citep[RSCG,][]{barton+96} identified using the friends-of-friends algorithm. 
However, only 5 of the 16 RSCGs are embedded in larger structures such as clusters.
Thus, Walker et al.'s samples could also be dominated by compact groups 
located in less dense regions.
We select 571 galaxies in the $\nbg\leq7$ compact groups to mimic the 
Walker et al.'s sample, and show their MIR color distribution 
in Figure \ref{comp_w13abhist_cluster}(b). 
The MIR color distribution of the 571 galaxies in the $\nbg\leq7$ compact groups
is not statistically different with the color distribution for 
the $0.4R_{200}\leq R<0.6R_{200}$ cluster galaxy sample
($P_{\rm KS}=0.444$ and $P_{\rm AD}=0.137$), which is consistent with 
the results of Walker et al. 
Therefore, the different results between this study and Walker et al. 
are probably from the use of different methods in identifying compact groups 
with and without using isolation criterion (see also Figure 10 of \citealt{sohn+16}).

One of the reasons why galaxy evolution is faster in compact groups than in clusters 
is likely because galaxy interactions are more frequent in compact groups 
than in clusters. 
The median velocity dispersion for our compact groups is  
$\sim$200 km s$^{-1}$, which is much smaller than the median value
for galaxy clusters, $\sim$800 km s$^{-1}$ \citep{rines+13}.
Furthermore, the mean size of compact groups is 30.7$\pm0.4$ kpc, which  
is much smaller than the typical virial radius of galaxy clusters, $\sim$1$-1.5$ Mpc
\citep{park+09,rines+13}. 
These characteristics make the interactions have a more significant effect 
in compact groups than in clusters.

\begin{figure}
\centering
\includegraphics[scale=0.7]{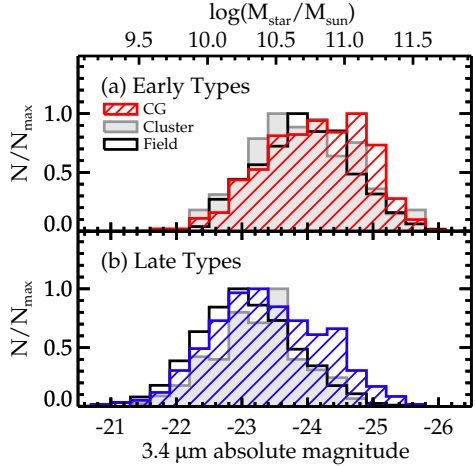}
\caption{Comparison of 3.4 $\micron$ absolute magnitude distributions 
between compact group (hatched histograms) and cluster galaxies (filled histograms).
The top panel is for early-type galaxies, while the bottom panel is 
for late-type galaxies.
}
\label{w1abs_env}
\end{figure}

In Figure \ref{w1abs_env}, we compare the 3.4 $\micron$ absolute 
magnitude ($M_{3.4}$) distribution of early- and late-type galaxies 
in compact groups, clusters, and the field. 
\citet{hwang+12b} showed that $M_{3.4}$ can be used as 
a proxy for stellar masses ($M_{\rm star}$) of galaxies.
We confirm the correlation between the $M_{3.4}$ and $M_{\rm star}$ 
using the compact group galaxies\footnote{log($M_{\rm star}/M_{\rm sun})=(-0.48\pm0.01)\times M_{3.4}-(0.76\pm0.20)$ with rms$=0.27$. }.
The stellar masses in this study are based on $H_0=70~\kms~{\rm Mpc}^{-1}$, 
and are calculated using the LePHARE\footnote{http://www.cfht.hawaii.edu/$\sim$arnouts/LEPHARE/lephare.html} spectral energy distribution fitting code 
\citep{arnouts+99,ilbert+06} with the SDSS $ugriz$ photometric data 
(see Section 2 of \citealt{sohn+16} for details). 
Table \ref{tab2} lists $M_{\rm star}$ and $M_{3.4}$ of compact group galaxies.

We find that among compact group early-type galaxies, the fraction of 
massive populations with $M_{3.4}<-24.5$ 
($M_{\rm star}\gtrsim10^{11} M_{\rm sun}$) is $34.9\%\pm1.8\%$, 
which is larger than the $24.6\%\pm2.1\%$ for the cluster early-type galaxy sample
and the $21.8\%\pm1.3\%$ for field early-type galaxy sample.
Similarly, among compact group late-type galaxies, the fraction of 
massive populations with $M_{3.4}<-23.5$ 
($M_{\rm star}\gtrsim10^{10.5} M_{\rm sun}$) is $44.4\%\pm2.5\%$, 
which is larger than the $32.9\%\pm3.2\%$ for the cluster late-type galaxy sample 
and the $27.0\%\pm0.8\%$ for the field late-type galaxy sample.
This result suggests that compact groups are an ideal environment for efficient
mass build-up of galaxies.

\subsection{Relation between Compact Group Members and Neighboring 
Galaxies}

\begin{figure*}
\centering
\includegraphics[scale=0.8]{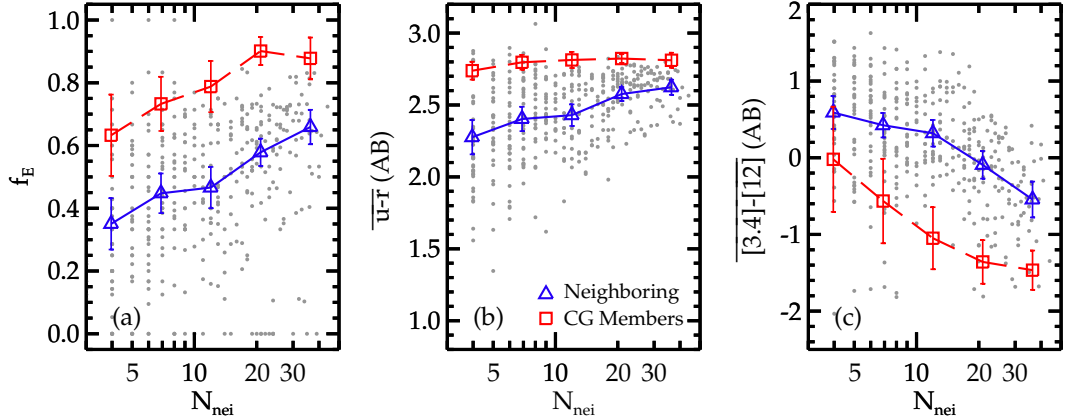}
\caption{$\nbg$ dependence of (a) $\efrac$, (b) $\overline{u-r}$, and 
(c) $\overline{[3.4]-[12]}$ for neighboring galaxies within the 
comoving cylinder around the compact groups, and their mean values (triangles) 
in $\nbg$ bins.
For comparison, we also plot the mean values of the three parameters 
for compact group member galaxies in $\nbg$ bins (squares).
Error bars represent $3\sigma$ deviation in a given $\nbg$ bin.
}
\label{nbg_bg_properties}
\end{figure*}

In Section 3.3, as shown in Figure \ref{w13abhist_nbg}, 
we find that compact groups have larger $\efrac$ 
and bluer MIR colors as their surrounding environments are denser (larger $\nbg$). 
To further investigate the effects of the surrounding environments on 
the physical properties of compact group member galaxies,
we compare the properties of neighboring galaxies around each compact group 
with those of the compact group member galaxies.
We calculate $\efrac$, mean $u-r$ ($\overline{u-r}$), and mean $[3.4]-[12]$ 
($\overline{[3.4]-[12]}$) for neighboring galaxies in the comoving cylinder of 
each compact group. 
Figure \ref{nbg_bg_properties} shows $\efrac$, $\overline{u-r}$, and 
$\overline{[3.4]-[12]}$ for neighboring galaxies as a function of $\nbg$. 
Triangles indicate the mean values of the three parameters in $\nbg$ bins.
As $\nbg$ increases, neighboring galaxies have larger $\efrac$, 
redder optical colors, and bluer MIR colors.
For comparison, we compute $\efrac$, $\overline{u-r}$, and $\overline{[3.4]-[12]}$ 
for member galaxies of each compact  group. Squares represent the 
mean values for the three parameters of compact group member galaxies.

We find that the two mean curves show a similar pattern in all three panels. 
The $\nbg$-dependence of $\efrac$, $\overline{u-r}$, and $\overline{[3.4]-[12]}$ 
for neighboring galaxies is consistent with the morphology-density or 
the SFR-density relation \citep[e.g.,][]{dressler80,lewis+02,park+09}.
However, it is interesting that the similar trends are also seen for compact group 
member galaxies. Since compact groups are very small in size (the mean size 
is $30.7\pm0.4$ kpc), their internal galaxy number densities are much higher 
than the surrounding densities. The mean internal density of compact groups, 
measured by the number of members divided by the sizes of the groups, is 
$1034\pm18$ Mpc$^{-2}$, which is much higher than the mean surrounding density,
$2.54\pm0.10$ Mpc$^{-2}$, from $\nbg$. 
Moreover, the internal galaxy number densities
of compact groups do not depend on the surrounding galaxy number densities. 
Thus, the $\nbg$-dependence of the properties of compact group member galaxies 
can not be explained by the morphology-density or SFR-density relations alone.

The similar $\nbg$-dependence for compact group member 
galaxies and their neighboring galaxies suggests that the properties of 
compact group galaxies are related to those of their neighboring galaxies.
If the physical properties of compact group galaxies are independent of their 
neighboring galaxies, the two mean curves would show different behaviors.
A plausible scenario would be that the neighboring galaxies are sources of 
member galaxies in compact groups. This supports the idea of \citet{diaferio+94} 
that compact groups replenish their members from surrounding environments, 
so that they do not disappear by mergers within a few Gyrs. 
Previous studies have supported this replenishment model by showing that 
many ($>50\%$) compact groups are associated with larger-scale galaxy structures
that may supply new members of compact groups. On the other hand, 
our finding is more direct observational evidence that the replenishment 
occurs regardless of the surrounding density of compact groups. 
 
An important point is that compact groups are not 
simply aggregates of captured neighboring galaxies. 
Figure \ref{nbg_bg_properties} shows that the two mean curves do not overlap, 
but differ systematically. This result suggests that there are additional environmental 
effects playing a critical role in morphology transformation and star formation 
quenching for compact group galaxies, which probably results from 
frequent interactions among member galaxies. 
This result also suggests that compact groups are the most suitable 
environment for the pre-processing \citep{zabludoff98}. 

\subsection{Hydrodynamic Interactions in Compact Groups}

We find that the MIR colors of late-type galaxies are bluer 
in the $\efrac>0$ compact groups than in the $\efrac=0$ compact groups 
(Figure \ref{w13_efrac_lgal}).
This $\efrac$ dependence is still found when $\nbg$ 
or $\sigma_{\rm CG}$ 
is fixed (see Figures \ref{w13abhist_vdisp_efrac} and 
\ref{w13abhist_nbg_efrac_lgal}), indicating that among the three environmental 
parameters,
$\efrac$ could be the most important parameter in determining the MIR colors 
of late-type member galaxies.
The $[3.4]-[12]$ colors of late-type galaxies are well correlated with specific 
SFRs \citep{donoso+12,hwang+12b}.
Thus, this finding suggests that star formation activity of late-type galaxies 
is suppressed more efficiently in the $\efrac>0$ compact groups than 
in the $\efrac=0$ compact groups. 
Moreover, the suppression of star formation activity is stronger 
in the $\efrac>0$ compact groups than in the cluster central ($R<0.5R_{200}$) regions
(Figure \ref{w13_efrac_lgal}).

\citet{bitsakis+10,bitsakis+11} showed that late-type galaxies in the $\efrac>0.25$ 
compact groups tend to have smaller specific SFRs than those in the $\efrac<0.25$ 
compact groups. Their result seems similar to our result. To verify their result 
with our sample, we divide our compact group sample into the $\efrac=0$, 
$0<\efrac\leq0.25$, and $\efrac>0.25$ compact groups. 
The large values ($>0.85$) of $P_{\rm KS}$ and $P_{\rm AD}$ cannot reject 
the null hypothesis for two MIR color distributions for late-type galaxies 
in the $0<\efrac\leq0.25$ compact groups and for late-type galaxies 
in the $\efrac>0.25$ compact groups.
However, the $P_{\rm KS}=0.056$ and $P_{\rm AD}=0.008$ reject the null hypothesis 
for two MIR color distributions for late-type galaxies in the $\efrac=0$ compact groups
and for late-type galaxies in the $0<\efrac\leq0.25$ compact groups 
at a significance $\gtrsim2\sigma$. 
These suggest that the $\efrac=0$ and $0<\efrac\leq0.25$ compact groups in 
our sample have statistically different MIR color distributions for
their late-type member galaxies, which supports again our finding that MIR colors 
of late-type galaxies are bluer in the $\efrac>0$ compact groups than 
in the $\efrac=0$ compact groups. The different $\efrac$ criterion 
between this study and Bitsakis et al. 
probably results from different compact group samples.

An important point here is that the star formation activity of late-type galaxies 
in compact groups is suppressed when the groups contain early-type members. 
This suppression of star formation activity cannot be explained by gravitational 
interactions among member galaxies alone. 
\citet{park+08} and \citet{pnc09} found that galaxy properties strongly depend on 
the distance and morphology of the nearest neighbor galaxy, using 
spectroscopic samples drawn from the SDSS data. They showed that 
when a galaxy is located within the virial radius of its nearest neighbor galaxy, 
the galaxy tends to have a morphological type similar to that of the neighbor galaxy. 
However, this phenomenon does not manifest 
if the distance to the neighbor is greater than the virial radius of the neighbor. 
They suggested that the phenomenon is due to hydrodynamic 
interactions, and that the effects of hydrodynamic interactions are significant to change the morphology and 
star formation activity of galaxies when the galaxies are located within the virial 
radii of their neighbor galaxies. 
The sizes of compact groups (the mean value is 30.7 kpc) are significantly
smaller than virial radii of member galaxies \citep[i.e., 430 kpc and 340 kpc 
for early- and late-type galaxies with $M_r=-20.77$,][]{park+09}. 
This implies that compact group member galaxies are already located 
within the virial radius of each other, and that they interact hydrodynamically 
each other.

\citet{park+09} focused on the case of galaxies located within the virial 
radii of massive galaxy clusters, and found that even in cluster environments
the hydrodynamic interactions with early-type neighbor galaxies are the main drivers
of star formation quenching of late-type galaxies. 
They also found that the effects of the cluster hot gas on the star formation quenching 
of galaxies at $R>0.1R_{200}$ is insignificant compared to the effects of hydrodynamic interactions with neighbor galaxies. Unlike cluster environments, hot gas of compact groups is not in a hydrostatic 
equilibrium state, and it is likely to be associated with the individual galaxies or 
brightest galaxies \citep{desjardins+13, desjardins+14}, suggesting that 
hydrodynamic effects from hot gas are insignificant in compact groups.
Therefore, we conclude that the suppressed star formation activity of late-type galaxies 
in the $\efrac>0$ compact groups likely results from hydrodynamic interactions 
with early-type member galaxies.

\section{Summary and Conclusions}\label{conclusions}

We study the MIR properties of galaxies in compact groups and 
their environmental dependence using a volume-limited sample 
of 670 compact groups and their 2175 member galaxies with 
$M_r<-19.77$ and $0.01<z<0.0741$, drawn from the catalog 
of \citet{sohn+16}. This catalog was derived from nearly complete redshift 
survey data, and was constructed by applying a friends-of-friends method
without applying Hickson's isolation criterion, which make the catalog 
include nearby compact groups and embedded compact groups in high-density regions
that were often excluded from previous catalogs. 
Using this unbiased compact group sample, we perform a study 
comparing the physical properties of galaxies in compact groups with 
those of galaxies in other environments such as galaxy clusters and the field. 
We use three environmental parameters including $\nbg$, $\efrac$, and 
$\sigma_{\rm CG}$ to represent the internal and external environment 
of the compact groups. Our key findings of this study are summarized as follows.

\begin{itemize}

\item[1.] The MIR colors of compact group early-type galaxies are on average
bluer than those of cluster ($R<0.5R_{200}$) early-type galaxies 
regardless of $\nbg$, $\efrac$, and $\sigma_{\rm}$.
This suggests that early-type galaxies in compact groups are on average older 
than those of cluster galaxies. 

\item[2.]  The MIR colors of the late-type galaxies in the $\efrac>0$ compact groups
tend to be bluer than those of cluster late-type galaxies.
These suggests that
the star formation activity of late-type galaxies is suppressed more efficiently 
in $\efrac>0$ compact groups than in clusters. 

\item[3.] As $\nbg$ increases, compact groups have larger $\efrac$, redder 
optical colors, and bluer MIR colors. These trends are also seen for 
neighboring galaxies around compact groups. 
Considering the extremely high galaxy number densities in compact group 
environments, this similar $\nbg$-dependence for compact group member galaxies 
and their neighboring galaxies is not well explained by the morphology- and 
SFR-density relations. This result can be explained by the scenario that 
neighboring galaxies are sources of member galaxies in compact groups, 
supporting the replenishment model suggested by \citet{diaferio+94}.

\item[4.] At a given $\nbg$, compact group members always have on average 
larger $\efrac$ and bluer MIR colors than neighboring galaxies.
This suggests that compact groups are not simply aggregates of captured 
neighboring galaxies, and that compact group environments play 
a critical role in accelerating morphology transformation and 
star formation quenching for the member galaxies.

\item[5.] In the $\efrac=0$ compact groups, the MIR colors of late-type galaxies 
are on average redder than those of cluster late-type galaxies. 
However, in the $\efrac>0$ compact groups, the MIR colors of late-type galaxies 
are on average bluer than those in cluster late-type galaxies.
This indicates that star formation quenching occurs more effectively 
in compact groups than in clusters when compact groups have early-type members. 
This suppressed star formation activity of late-type galaxies likely results from 
hydrodynamic interactions with early-type member galaxies.

\end{itemize}

Our results suggest that galaxy evolution is faster in compact groups than 
in other environments, and that compact groups are likely to be the best 
place for the pre-processing \citep{zabludoff98}.

\acknowledgments
We thank the anonymous referee for useful and constructive comments 
that helped us to improve the manuscript.
We also thank Brian S. Cho for helping the English in the manuscript improved.
G.H.L. acknowledges the support by the National Research Foundation of Korea (NRF) Grant funded 
by the Korean Government (NRF-2012-Fostering Core Leaders of the Future Basic Science Program).
M.G.L was supported by the NRF grant funded by the Korea Government (MEST) (No. 2012R1A4A1028713).

\end{document}